\documentclass{aa}  

\usepackage{float} 

\usepackage{textcomp}
\usepackage[utf8]{inputenc}

\usepackage{amsmath, amssymb, amsfonts, amscd}
\usepackage{mathtools} 
\usepackage{array} 
\usepackage{cases} 
\usepackage{empheq} 
\usepackage{multirow} 
\usepackage{txfonts} 

\usepackage{graphicx} 
\usepackage{wrapfig} 

\usepackage{natbib}
 
\begin{document}

\title{The impact of rotation on turbulent tidal friction\\ in stellar and planetary convective regions}

\author{
S. Mathis\inst{1,2} \and
P. Auclair-Desrotour\inst{3,1} \and
M. Guenel\inst{1} \and
{F. Gallet}\inst{4} \and
C. Le Poncin-Lafitte\inst{5}
}

\institute{Laboratoire AIM Paris-Saclay, CEA/DSM - CNRS - Universit\'e Paris Diderot, IRFU/SAp Centre de Saclay, F-91191 Gif-sur-Yvette Cedex, France
\and LESIA, Observatoire de Paris, PSL Research University, CNRS, Sorbonne Universit\'es, UPMC Univ. Paris 06, Univ. Paris Diderot, Sorbonne Paris Cit\'e, 5 place Jules Janssen, 92195 Meudon, France
\and IMCCE, Observatoire de Paris, UMR 8028 du CNRS, UPMC, 77 Av. Denfert-Rochereau, 75014 Paris, France
\and Department of Astronomy, University of Geneva, Chemin des Maillettes 51, 1290 Versoix, Switzerland
\and SYRTE, Observatoire de Paris, PSL Research University, CNRS, Sorbonne Universit\'es, UPMC Univ. Paris 06, LNE, 61 avenue de l'Observatoire, 75014 Paris, France\\
\email{stephane.mathis@cea.fr;pierre.auclair-desrotour@obspm.fr; mathieu.guenel@cea.fr;\\florian.gallet@unige.ch;christophe.leponcin@obspm.fr} 
}

\date{Received ... / accepted ...}

\abstract 
{Turbulent friction in convective regions in stars and planets is one of the key physical mechanisms that drive the dissipation of the kinetic energy of tidal flows in their interiors and the evolution of their systems. This friction acts both on the equilibrium/non-wave like tide and on tidal inertial waves in these layers.}
{It is thus necessary to obtain a robust prescription for this friction. In the current state-of-the-art, it is modeled by a turbulent eddy-viscosity coefficient, based on mixing-length theory, applied on velocities of tides. However, none of the current prescriptions take into account the action of rotation that can strongly affects turbulent convection. Therefore, a new prescription that takes this latter into account must be derived.}
{We use theoretical scaling laws for convective velocities and characteristic lengthscales in rotating stars and planets that have been recently confirmed by 3-D high-resolution nonlinear Cartesian numerical simulations to derive the new prescription. A corresponding local model of tidal waves is used to understand the consequences for the linear tidal dissipation. Finally, new grids of rotating stellar models and published values of planetary convective Rossby numbers are used to discuss astrophysical consequences.}
{The action of rotation on convection deeply modifies the turbulent friction applied on tides. In the regime of rapid rotation (with a convective Rossby number below 0.25), the eddy-viscosity may be decreased by several ordres of magnitude. It may lead to a loss of efficiency of the viscous dissipation of the equilibrium tide and to a more efficient complex and resonant dissipation of tidal inertial waves in the bulk of convective regions.}
{To understand the complete evolution of planetary systems, tidal friction in rapid rotators such as young low-mass stars, giant and Earth-like planets must be evaluated. Therefore, it is necessary to have a completely coupled treatment of the tidal evolution of star-planet systems and multiple stars and of the rotational evolution of their components with a coherent treatment of the variations of tidal flows and of their dissipation as a function of rotation.}


\titlerunning{The impact of rotation on turbulent tidal friction in stellar and planetary convective regions}
\authorrunning{Mathis et al.}

\maketitle


\section{Introduction and context}

Tidal friction is one of the mechanisms that drives the evolution of star-planet and planet-moon systems \citep[e.g.][]{Hut1981,Letal2012,Bolmontetal2012}. It shapes their orbital architecture and the rotational dynamics of each of their components. Its properties strongly depend on their internal structure and dynamics \citep[e.g.][]{GS1966,MathisRemus2013,Ogilvie2014}. Indeed, tidal dissipation that converts the kinetic energy of tidal flows into heat in stars and fluid planetary layers strongly differs from the one in rocky/icy regions \citep{EL2007,Aetal2014}. Actually, the variation of tidal dissipation in fluids as a function of the forcing frequency is strongly resonant \citep{OL2004,OL2007,ADMLP2015}. These resonances are due to the excitation of low-frequency inertial waves in convective layers and of gravito-inertial waves in stably-stratified regions. Their properties are function of rotation, stratification, and viscous and thermal diffusivities \citep{Zahn1975,OL2004,OL2007,ADMLP2015}. Finally, more and more observational constrains become available in the Solar and extrasolar systems \citep[see e.g.][and reviews done in Ogilvie 2014 and Auclair-Desrotour et al. 2015]{Laineyetal2009,Laineyetal2012,Winnetal2010,Albrechtetal2012,VR2014a,VR2014b}.\\

In this context, tidal friction in the turbulent convective envelopes of low-mass stars (from M to F stellar types) and giant planets and the cores of telluric planets must be carefully evaluated. In the present state-of-the-art, the turbulent friction acting on tidal flows in these regions is modeled thanks to an effective turbulent viscosity coefficient. This corresponds to the assumptions that we have a scale-separation between tidal and turbulent convective flows, that turbulence is close to be isotropic and that the friction can be described through a viscous force \citep{Zahn1966}. The properties of the turbulent viscosity thus describe the effective efficiency of the couplings between turbulence and tidal flows. Therefore, it depends on the frequency of the forcing and on the dynamical parameters that impact stellar and planetary convection \citep{Zahn1966,GK1977,Zahn1989,GoodmanOh1997,Penevetal2007,OL2012}. 

Among them, rotation is one of the parameters that must be taken into account. Indeed, the Coriolis acceleration strongly affects the dynamics of turbulent convective flows \citep[e.g.][]{Brownetal2008,Julienetal2012,Barkeretal2014}. In this framework, the rotation of stars and planets can strongly vary along the evolution of planetary systems \citep[e.g.][for low-mass stars]{Bouvier2008,GalletBouvier2013,GalletBouvier2015,Amardetal2016}. Therefore, it is absolutely necessary to get a robust prescription for the turbulent friction applied on tidal waves by rotating convection in stellar and planetary interiors as a function of their angular velocity. To reach this objective, properties of rotating turbulent flows such as their characteristic velocities and length scales must be known if we wish to model this friction using the mixing-length framework \citep{Zahn1966}. In this context, the work by \cite{Stevenson1979} is particularly interesting since he derived them in the asymptotic regimes of slow and rapid rotation. Moreover, these asymptotic scaling laws have been now confirmed by \cite{Barkeretal2014} in the regime of rapid rotation using high-resolution non-linear 3-D Cartesian simulations of turbulent convection.

In this work, we thus propose a new prescription for tidal friction in rotating turbulent convective stellar and planetary zones that takes rotation into account using the results obtained by \cite{Stevenson1979} and \cite {Barkeretal2014}. First, in section 2, we recall the state-of-the-art prescriptions that do not take rotation into account. In section 3, we propose our new prescription for the turbulent friction applied by rotating convective flows. In section 4, we discuss consequences for the viscous dissipation of tidal flows in rotating stellar and planetary convection zones and corresponding scaling laws \citep[][]{ADMLP2015}. In section 5, we examine consequences for tidal dissipation in convective regions in low-mass stars during the Pre-Main-Sequence (hereafter PMS) and the Main Sequence (hereafter MS) and in planets. Finally, we conclude and present the perspectives of this work for the evolution of planetary systems.

\section{State of the art}
\label{sec:NR}
The first study of the friction applied by turbulent convection on tidal flows was achieved by \cite{Zahn1966} in the case of binary stars \citep[see also][]{Zahn1989}. In his work, he examined the coupling between turbulence and the large-scale equilibrium/non wave-like tide induced by the hydrostatic adjustment of the star due to the tidal perturber \citep[e.g.][]{Zahn1966,RMZ2012,Ogilvie2013}. His approach was based on three main assumptions. First, he assumed a scale-separation between tidal and turbulent convective flows. Second, he assumed that the friction applied by turbulence can be described thanks to a viscous force involving an eddy-viscosity $\nu_{\rm T}$. This implies the third assumption of an isotropic turbulence. Finally, the characteristic velocity and length scale of turbulent convection, respectively $V_{\rm c}$ and $L_{\rm c}$, were described using the mixing-length theory. We have
\begin{equation} 
V_{\rm c}\left(\Omega=0\right)\approx\left(\frac{L_{\rm b}}{{\overline\rho}_{\rm CZ}R^2}\right)^{1/3}\quad\hbox{and}\quad L_{\rm c}\left(\Omega=0\right)\approx\alpha H_{\rm p}, 
\end{equation}
where $L_{\rm b}$, $R$, ${\overline\rho}_{\rm CZ}$, $\alpha$, $H_{\rm p}$ and $\Omega$ are the body luminosity and radius, the mean density in the studied convective region, the free mixing-length parameter, the pressure height-scale and the angular velocity, respectively \citep[][]{Brun2014}. In this framework, he derived the following prescription for the eddy-viscosity:
\begin{equation}
\nu_{\rm T; NR}=\frac{1}{3}V_{\rm c}\left(\Omega=0\right) L_{\rm c}\left(\Omega=0\right)\,f\left(\frac{P_{\rm tide}}{P_{\rm c}}\right).
\label{nuNR}
\end{equation}
In this expression, NR stands for {\it Non-Rotating convection} and $f$ is a function that describes the loss of efficiency of tidal friction in convective regions in the case of rapid tide when $P_{\rm tide}\!<\!\!<\!P_{\rm c}$, $P_{\rm tide}$ and $P_{\rm c}=L_{\rm c}/V_{\rm c}$ being respectively the tidal period and the characteristic convective turn-over time. Two expressions have been proposed for $f$ in the literature by \cite{Zahn1966} and \cite{GK1977}: \cite{Zahn1966} proposed a linear attenuation with $f\propto\left(P_{\rm tide}/P_{\rm c}\right)$ and \cite{GK1977} a quadratic one, i.e. $f\propto\left(P_{\rm tide}/P_{\rm c}\right)^{2}$. These prescriptions have been examined both by theoretical work \citep{GoodmanOh1997} and by local high-resolution 3D numerical simulations \citep[][]{Penevetal2007,OL2012}. If numerical simulations computed by \cite{Penevetal2007} seems to confirm the linar attenuation proposed by \cite{Zahn1966}, those by \cite{OL2012} are in favor of the quadratic one, leading to a situation where the prescription that must be used for $f$ is still debated.\\

Moreover, as pointed above, (rapid) rotation strongly affects turbulent convective flows \citep[e.g.][]{Brownetal2008,Julienetal2012,Barkeretal2014}. First, the Coriolis acceleration stabilizes the flow leading to a shift of the threshold of the convective instability \citep[][]{Chandrasekhar1953}. Next, the efficiency of the heat transport and the turbulent energy cascade are inhibited \citep[e.g.][]{Senetal2012,Kingetal2012,Kingetal2013,Barkeretal2014}. Finally, $V_{\rm c}$, $L_{\rm c}$, and as a consequence $\nu_{\rm T}$ vary with rotation. In the present state-of-the-art, we are thus in a situation where the action of the Coriolis acceleration is taken into account in the physical description of tidal flows \citep[e.g.][]{OL2004,OL2007,RMZ2012,ADMLP2015} while it is ignored in the description of the turbulent friction. This should be improved since the angular velocity of celestial bodies can vary by several ordres of magnitude along their evolution. Therefore, the turbulent convective friction must also be described as a function of the rotation rate ($\Omega$).

\section{Prescription for the friction applied by turbulent rotating convection}

\subsection{Modelling and assumptions}

\begin{figure}[t!]
\begin{center}
\includegraphics[width=0.9\linewidth]{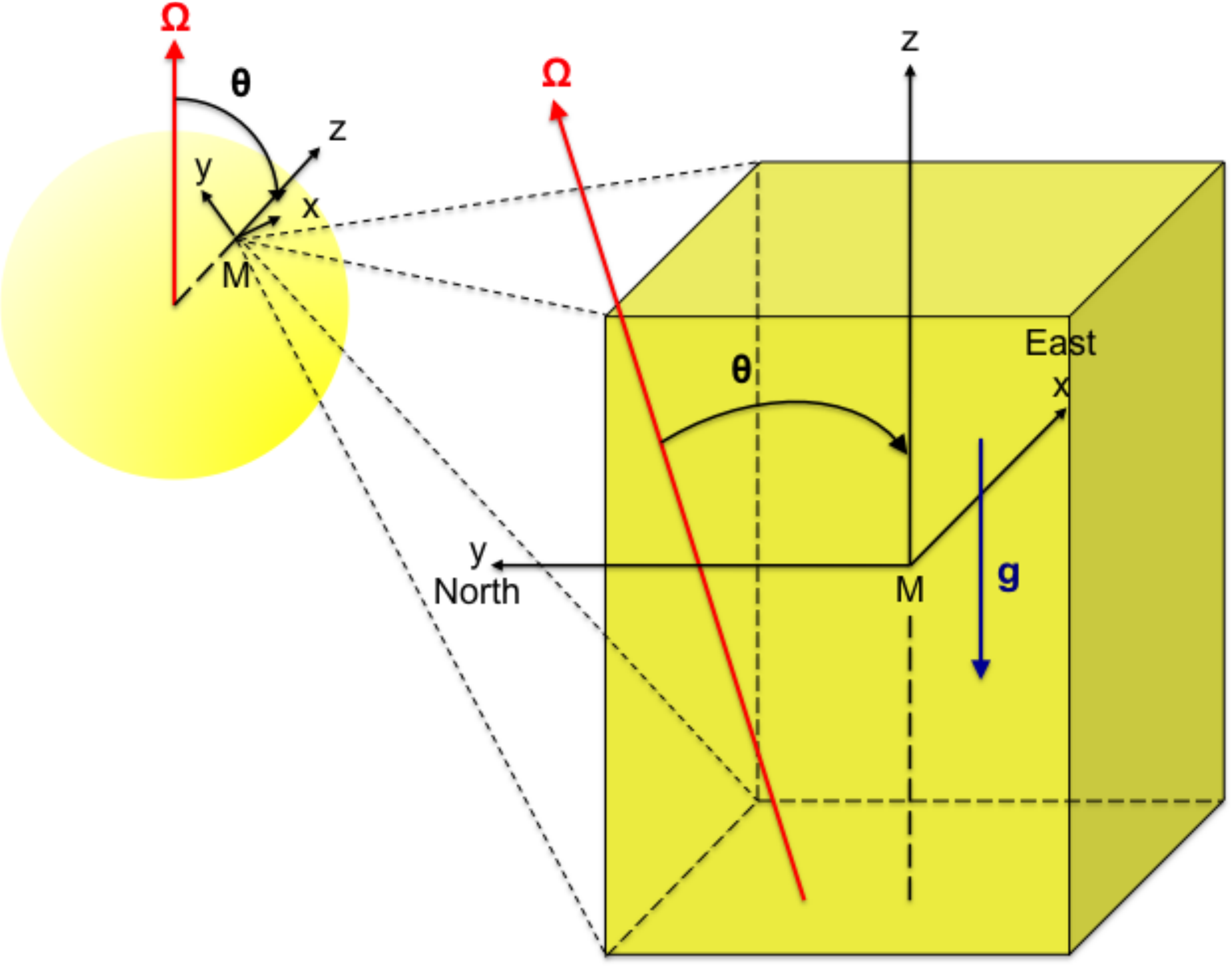}
\end{center} 
\caption{The local Cartesian model. M is the origin of the set-up. The east, north and gravity ($\vec g$) directions are along the $x$, $y$ and $z$ axis, respectively. The angle $\theta$ is the inclination of the rotation axis with respect to gravity.}
\label{fig:Repere}
\end{figure}

To study the modification of the turbulent friction applied on tidal flows in rotating stellar and planetary convective regions, we will use theoretical results first derived by \cite{Stevenson1979} and confirmed in the rapid rotation regime by high-resolution numerical Cartesian simulations computed by \cite{Barkeretal2014}. We thus choose to consider a local Cartesian set-up with a box centered around a point $M$ in a rotating convective zone (see Fig. \ref{fig:Repere}) with an angular velocity $\Omega$; $\left(M,x,y,z\right)$ is the associated reference frame. The box has a characteristic length $L$ and is assumed to have a homogeneous density $\rho$. Its vertical axis, which is aligned with the gravity ${\vec g}$, is inclined with an angle $\theta$ with respect to the rotation axis. The convective turbulent flow has characteristic velocity $V_{\rm c}$ and length scale $L_{\rm c}$ to which we associate the kinematic eddy-viscosity $\nu_{\rm T}$ (see Eqs. \ref{nuNR} and \ref{nuR}) describing the turbulent friction.

Next, we introduce the control parameters of the system:
\begin{itemize}
\item the convective Rossby number defined as in \cite{Stevenson1979}
\begin{equation}
R_{\rm o}^{\rm c}= \left(\displaystyle \frac{V_{\rm c}\left(\Omega=0\right)}{2\Omega L_{\rm c}\left(\Omega=0\right)\vert\cos\theta\vert}\right)=\displaystyle{\frac{P_{\Omega}}{P_{\rm c}\left(\Omega=0\right)}},
\label{defRoc}
\end{equation}
where we introduce the dynamical time $P_{\Omega}=\displaystyle{\frac{1}{2\Omega\vert\cos\theta\vert}}$ and we recall the definition of the characteristic convective turn-over time $P_{\rm c}=L_{\rm c}/V_{\rm c}$; $R_{\rm o}^{\rm c}\!<\!\!<\!1$ and $R_{\rm o}^{\rm c}\!>\!\!>\!1$ correspond to rapid and slow rotation regimes, respectively;
\item the Ekman number
\begin{equation}
E = \displaystyle \frac{2 \pi^2 \nu_{\rm T}}{ \Omega L^2},
\label{Eckman}
\end{equation}
which compares the respective strength of the viscous force and of the Coriolis acceleration \citep[see also][]{ADMLP2015}. 
\end{itemize}

\subsection{The new eddy-viscosity prescription}

As in previous works, which do not take into account the action of rotation on convection (see Sec. \ref{sec:NR}), we assume that: i) we have a scale-separation between turbulent convective flows and tidal velocities and ii) the turbulent friction on this latter can be modeled through a viscous force involving an eddy-viscosity coefficient. 

\begin{figure}[t!]
\begin{center}
\includegraphics[width=0.9\linewidth]{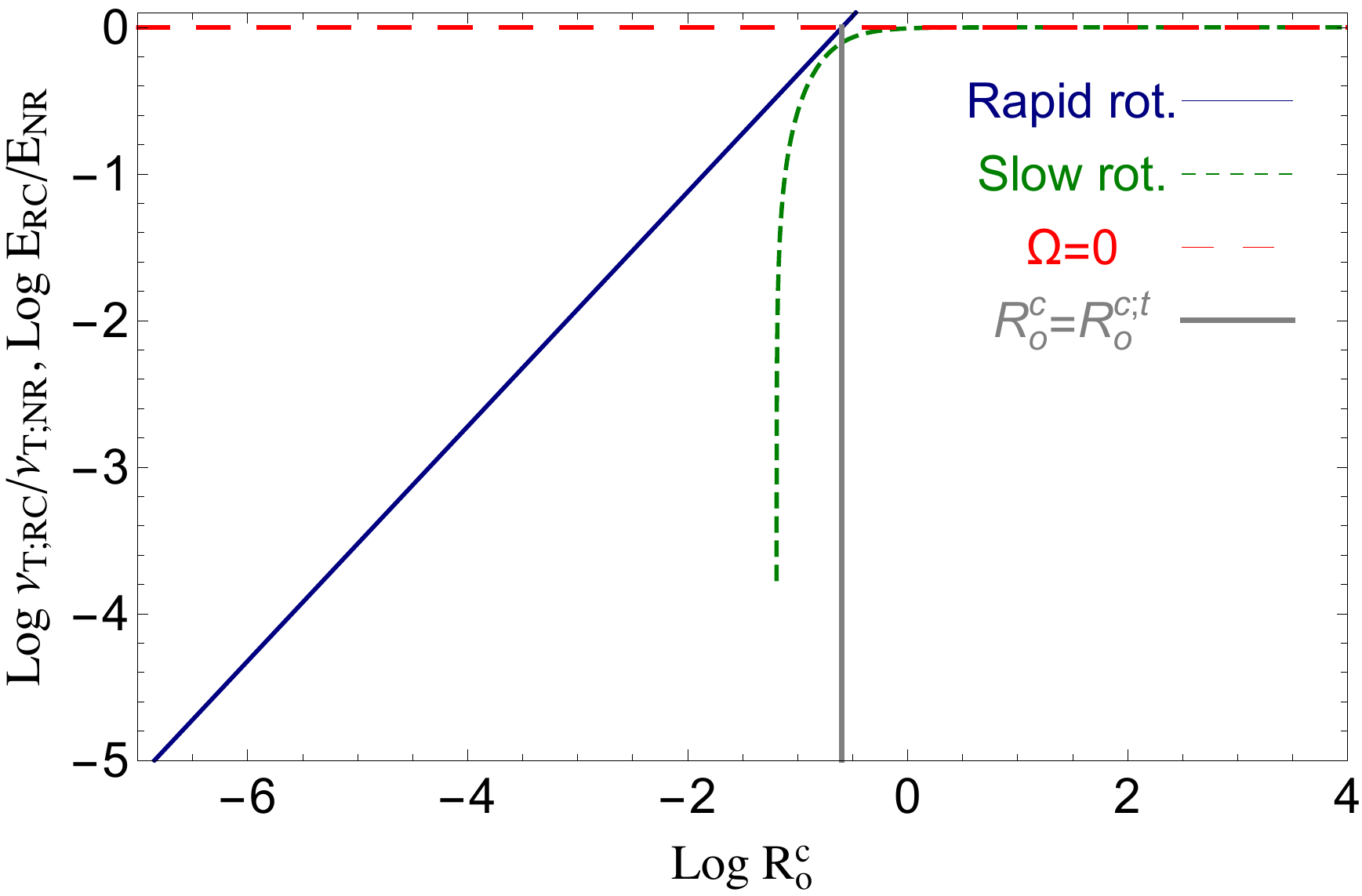}
\end{center} 
\caption{{The logarithm of the ratio $\nu_{\rm T; RC}/\nu_{\rm T;NR}$ (and $E_{\rm RC}/E_{\rm NR}$) as a function of $\log R_{\rm o}^{\rm c}$}. The small-dashed green and solid blue lines correspond to the slow- and rapid-rotation asymptotic regimes, respectively. The vertical grey solid line corresponds to the transition Rossby number $R_{\rm o}^{\rm c;t}\approx0.25$ between these two asymptotic regimes. The red long-dashed line corresponds to the non-rotating case.}
\label{fig:TurbVisc}
\end{figure}

To derive this coefficient, we have to know the variation of $V_{\rm c}$ and $L_{\rm c}$ as a function of $\Omega$ and to verify that the mixing-length approach that is generally used in stellar and planetary models can be assumed in our context. Actually, in presence of (rapid) rotation, convective turbulence becomes highly anisotropic because of the action of the Coriolis acceleration \citep[e.g.][]{Julienetal2012,Senetal2012,Kingetal2012,Kingetal2013} and one must verify that a simplified mixing-length approach can be assumed as a first step. In this framework, this is the great interest of the work by \cite{Barkeretal2014}. It demonstrated that scaling laws obtained for $V_{\rm c}$ and $l_{\rm c}$ as a function of $R_{\rm o}^{\rm c}$ by \cite{Stevenson1979} using such a mixing-length formalism is robust and verified when computing high-resolution Cartesian numerical simulations of rapidly rotating turbulent convective flows in a set-up corresponding to the one studied here. 

We can thus generalize the prescription proposed in Eq. (\ref{nuNR}) to the rotating case by writing
\begin{equation}
\nu_{\rm T; RC}=\frac{1}{3}V_{\rm c}\left(R_{\rm o}^{\rm c}\right) L_{\rm c}\left(R_{\rm o}^{\rm c}\right)\,f\left(\frac{P_{\rm tide}}{P_{\rm c}}\right),
\label{nuR}
\end{equation}
where RC stands for {\it Rotating Convection}.
To get $V_{\rm c}\left(R_{\rm o}^{\rm c}\right)/V_{\rm c}\left(\Omega=0\right)$ and $L_{\rm c}\left(R_{\rm o}^{\rm c}\right)/L_{\rm c}\left(\Omega=0\right)$, we use the scaling laws that have been derived by \cite{Stevenson1979} and verified by \cite{Barkeretal2014} in the rapidly rotating regime:
\begin{itemize}
\item in the {\it slow rotation} regime, we have
\begin{equation}
\frac{V_{\rm c}\left(R_{\rm o}^{\rm c}\right)}{V_{\rm c}\left(\Omega=0\right)}=\left(1-\frac{1}{242\left(R_{\rm o}^{\rm c}\right)^{2}}\right)
\end{equation}
and
\begin{equation}
\frac{L_{\rm c}\left(R_{\rm o}^{\rm c}\right)}{L_{\rm c}\left(\Omega=0\right)}=\left(1+\frac{1}{82\left(R_{\rm o}^{\rm c}\right)^{2}}\right)^{-1};
\end{equation}
\item in the {\it rapid rotation} regime, we have
\begin{equation}
\frac{V_{\rm c}\left(R_{\rm o}^{\rm c}\right)}{V_{\rm c}\left(\Omega=0\right)}=1.5(R_{\rm o}^{\rm c})^{1/5}\quad\hbox{and}\quad\frac{L_{\rm c}\left(R_{\rm o}^{\rm c}\right)}{L_{\rm c}\left(\Omega=0\right)}=2(R_{\rm o}^{\rm c})^{3/5}.
\label{RR}
\end{equation}
\end{itemize}

In Fig. \ref{fig:TurbVisc}, we plot $\log\left(\nu_{\rm T; RC}/\nu_{\rm T;NR}\right)$ and the corresponding ratio for the Ekman number (see Eq. \ref{Eckman}) as a function of $\log R_{\rm o}^{\rm c}$. We have define a first Ekman number computed with the turbulent viscosity prescription where the modification of turbulent friction by rotation is ignored
\begin{equation}
E_{\rm NR}=\frac{2\pi^2\nu_{\rm T;NR}}{\Omega L^2}
\label{def:ENR}
\end{equation}
and a second one where it is taken into account
\begin{equation}
E_{\rm RC}=\frac{2\pi^2\nu_{\rm T;RC}}{\Omega L^2}.
\label{def:ERC}
\end{equation}
The small-dashed green and solid blue lines correspond to the slow- and rapid-rotation asymptotic regimes respectively (the red long-dashed line corresponding to the non-rotating case). The vertical grey solid line corresponds to the transition Rossby number $R_{\rm o}^{\rm c;t}\approx0.25$ between these two asymptotic regimes. In the regime of rapidly rotating convective flows ($R_{\rm o}^{\rm c}\!<\!\!<\!1$), the turbulent friction decreases by several ordres of magnitude with a scaling $\nu_{\rm T; RC}/\nu_{\rm T;NR}\propto\left(R_{\rm o}^{\rm c}\right)^{4/5}\propto\Omega^{-4/5}$. It can be understood coming back on the action of (rapid) rotation on the convective instability and turbulence (see the discussion at the end of Sec. \ref{sec:NR}). Indeed, the Coriolis acceleration tends to stabilize the flow and thus the degree of turbulence decreases with increasing rotation as well as the corresponding turbulent friction and eddy-viscosity. 

Consequences for the viscous dissipation of the kinetic energy of tidal flows in rotating stellar and planetary convective regions have now to be examined.

\section{Consequences for tidal dissipation}

\subsection{A local model to quantify tidal dissipation}
\label{sec:scalinglaws}

The linear response of planetary and stellar rotating convection zones to tidal perturbations is constituted by the superposition of an equilibrium/non-wave like tide displacement and of tidally excited inertial waves, the dynamical tide \citep[e.g.][]{Zahn1966,OL2004,OL2007,RMZ2012,Ogilvie2013}. The restoring force of inertial waves is the Coriolis acceleration. Because of their dispersion relation $\sigma=\pm2\Omega\,k_{z}/\vert{\vec k}\vert$, where $\sigma$ is their frequency and $\vec k$ their wave number, they propagate only if $\sigma\in\left[-2\Omega,2\Omega\right]$. 

To understand the impact of rapid rotation on the turbulent friction derived in the previous section on the equilibrium and dynamical tides, we now consider the linear response of the Cartesian set-up studied here (cf. Fig. \ref{fig:Repere}) to a periodic tidal forcing. As a first step, we thus neglect the non-linear interactions between tidal inertial waves and those with turbulent convective flows \citep[see e.g.][]{Galtier2003,Senetal2012,Favieretal2014,diLeonietal2014,Campagneetal2015}. We follow the reduced and local analytical approach introduced by \cite{OL2004} in the appendix of their paper and generalized by \cite{ADMLP2015} to understand tidal dissipation in convective regions with taking into account here the inclination angle $\theta$. In this framework, the velocity field of the tide $\vec u$ excited by the tidal periodic volumetric forcing $\vec F$ per unit-mass is governed by the linearized momentum and continuity equations\footnote{In this work the action of thermal diffusivity is ignored.}:
\begin{equation}
\partial_{t}{\vec u}+2\vec\Omega\times\vec u=-\nabla\Pi+\nu\nabla^{2}\vec u+\vec F\quad\hbox{and}\quad\vec\nabla\cdot\vec u={\vec 0}\,,
\end{equation}
where $\nu$ is the (effective turbulent) viscosity and $\Pi=P/\rho$ with $P$ and $\rho$ being the pressure and density respectively. We follow \cite{OL2004} and \cite{ADMLP2015} by expanding $\vec u$, $\Pi$, and $\vec f=\vec F/2\Omega$ as Fourier series in time and space
\begin{equation}
\begin{array}{c c c}
   u_x = \Re \left[ u(X,Z) e^{-i \omega T}  \right], & u_y = \Re \left[ v(X,Z) e^{-i \omega T}  \right], \\ 
   u_z = \Re \left[ w(X,Z) e^{-i \omega T}  \right], & \Pi = \Re \left[ \psi (X,Z) e^{-i \omega T}  \right], \\
   f_x = \Re \left[ f(X,Z) e^{-i \omega T}  \right], & f_y = \Re \left[ g (X,Z) e^{-i \omega T}  \right], \\
   f_z = \Re \left[ h(X,Z) e^{-i \omega T}  \right],\\
\end{array}
\end{equation}
with
\begin{equation}
\begin{array}{c c }
   u = \displaystyle \sum_{m,n} u_{mn} e^{i 2 \pi \left( mX + n Z  \right) }, & v = \displaystyle \sum_{m,n} v_{mn} e^{i 2 \pi \left( mX + n Z  \right) },\\
   \vspace{0.1mm}\\
   w = \displaystyle \sum_{m,n} w_{mn} e^{i 2 \pi \left( mX + n Z  \right) }, & \psi = \displaystyle \sum_{m,n} \psi_{mn} e^{i 2 \pi \left( mX + n Z  \right) },\\
   \vspace{0.1mm}\\
   f = \displaystyle \sum_{m,n} f_{mn} e^{i 2 \pi \left( mX + n Z  \right) }, & g = \displaystyle \sum_{m,n} g_{mn} e^{i 2 \pi \left( mX + n Z  \right) },\\
   \vspace{0.1mm}\\
   h = \displaystyle \sum_{m,n} h_{mn} e^{i 2 \pi \left( mX + n Z  \right) }.\\
\end{array}
\end{equation}
We have introduced the normalized space coordinates\footnote{As in \cite{OL2004} and \cite{ADMLP2015}, we focus to solutions, which depend only on $X$ and $Z$, since adding the third coordinate $Y$ will not modify the qualitative behavior of the system.} $X=x/L$ and $Z=z/L$, horizontal and vertical wave numbers $m$ and $n$, time $T=2\Omega\,t$, and tidal frequency $\omega=\left(s{\widetilde n}-M\Omega\right)/\left(2\Omega\right)$; ${\widetilde n}$ is the mean orbital motion, $s\in{\pmb Z}$ and $M=mL/\left(r\sin\theta\right)=\left\{1,2\right\}$ ($r$ is the radial spherical coordinate of M). The boundary conditions are periodic in the two directions that corresponds to normal modes excited by tides \cite[e.g.][]{Wu2005,BO2015}. It would be also possible to tackle the case of singular modes leading to inertial wave attractors by imposing rigid boundary conditions to our tilted Cartesian box \citep{O2005,JO2014}.

From the momentum and the continuity equations, we get the following system:
\begin{equation}
\left\{
\begin{array}{rcl}
\!\!  - i\omega u_{mn}  - \cos \theta v_{mn} + \sin \theta w_{mn} \!\!\!\!\! & = & \!\!\!\!\! - i m \Lambda \psi_{mn} - E \left( m^2 + n^2 \right)u_{mn} \\
     & & + f_{mn}\\
   \vspace{0.1mm}\\
  - i\omega v_{mn} + \cos \theta u_{mn} \!\!\!\!\! & = & \!\!\!\!\! - E \left( m^2 + n^2 \right)v_{mn} + g_{mn}\\
  \vspace{0.1mm}\\
  - i \omega w_{mn} - \sin \theta u_{mn} \!\!\!\!\! & = & \!\!\!\!\! - i n \Lambda \psi_{mn} - E \left( m^2 + n^2 \right)w_{mn} \\
  & & + h_{mn} \\
  \vspace{0.1mm}\\
  m u_{mn} + n w_{mn} \!\!\!\!\! & = & \!\!\!\!\! 0
\end{array}
\right. ,
\label{systeme_eq}
\end{equation}
where $\Lambda=1/\left(2\Omega\,L\right)$. Following \cite{ADMLP2015}, we solve it analytically. This allows us to derive each Fourier coefficient of the velocity field
\begin{equation}
\left\{
\begin{array}{ccl}
  u_{mn} & = & \displaystyle  n \frac{i \tilde{\omega} \left( n f_{mn} - m h_{mn} \right) - n \cos \theta g_{mn}  }{ \left( m^2 + n^2 \right) \tilde{\omega}^2 - n^2 \cos^2 \theta}\\
  \vspace*{0,1mm}\\
  v_{mn} & = & \displaystyle\frac{n \cos \theta \left( n f_{mn} - m h_{mn}  \right) + i \left[ \left( m^2 + n^2 \right) \tilde{\omega}\right] g_{mn} }{\left( m^2 + n^2 \right) \tilde{\omega}^2 - n^2 \cos^2 \theta}\\
  \vspace*{0,1mm}\\
  w_{mn} & = & - m \displaystyle \frac{i \tilde{\omega} \left( n f_{mn} - m h_{mn} \right) - n \cos \theta g_{mn} }{\left( m^2 + n^2 \right) \tilde{\omega}^2 - n^2 \cos^2 \theta}\\
\end{array}
\right. ,
\label{v:sol}
\end{equation}
with $\tilde{\omega} = \omega + i E\left( m^2 + n^2 \right)$, and to compute the viscous dissipation per unit mass of the kinetic energy of tidal flows \citep{ADMLP2015}
\begin{eqnarray}
\lefteqn{{\mathcal D} = \frac{1}{L^2}\int_0^1 \int_0^1 \left\langle - \textbf{u} \cdot \nu \nabla_{X,Z}^2 \textbf{u} \right\rangle {\rm d}X{\rm d}Z}\nonumber\\
& = & \frac{2 \pi^2 \nu}{ L^2}  \sum_{ (m,n) \in \mathbb{Z^*}^2 } \left( m^2 + n^2 \right) \left( \left| u_{mn}^2 \right| + \left| v_{mn}^2 \right| + \left| w_{mn}^2 \right| \right),
\end{eqnarray}
where $\nabla_{X,Z}^{2}=1/L^2\nabla^{2}$ and $\left<\cdot\!\cdot\!\cdot\right>$ is the average in time, and the corresponding energy dissipated per rotation period
\begin{equation}
\zeta = \frac{2 \pi}{\Omega}{\mathcal D} = 2 \pi E \sum_{ (m,n) \in \mathbb{Z^*}^2 } \left( m^2 + n^2 \right) \left( \left| u_{mn}^2 \right| + \left| v_{mn}^2 \right| + \left| w_{mn}^2 \right| \right).
\label{dissipation}
\end{equation}

\begin{figure}[t!]
\begin{center}
\includegraphics[width=0.85\linewidth]{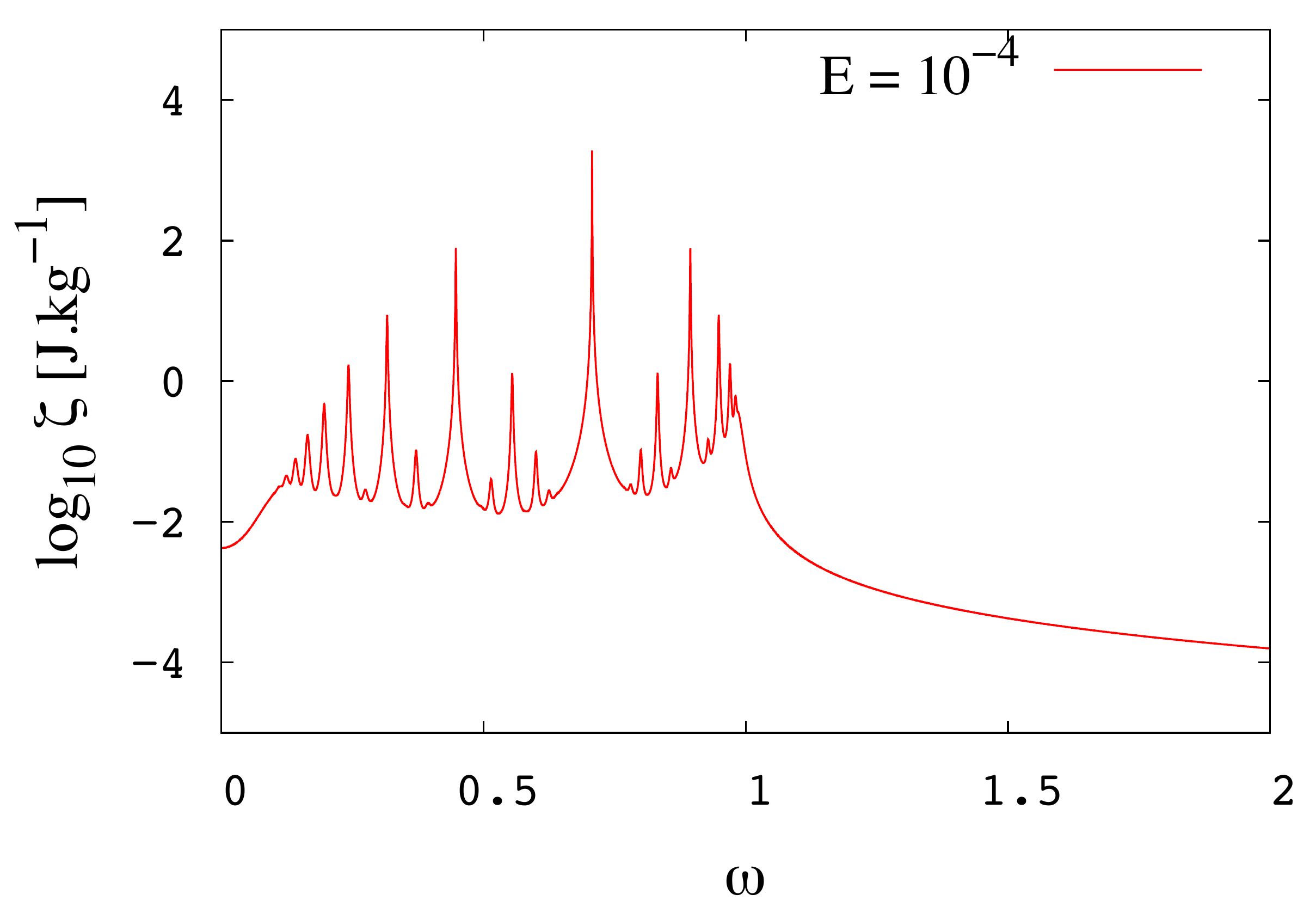}
\includegraphics[width=0.9\linewidth]{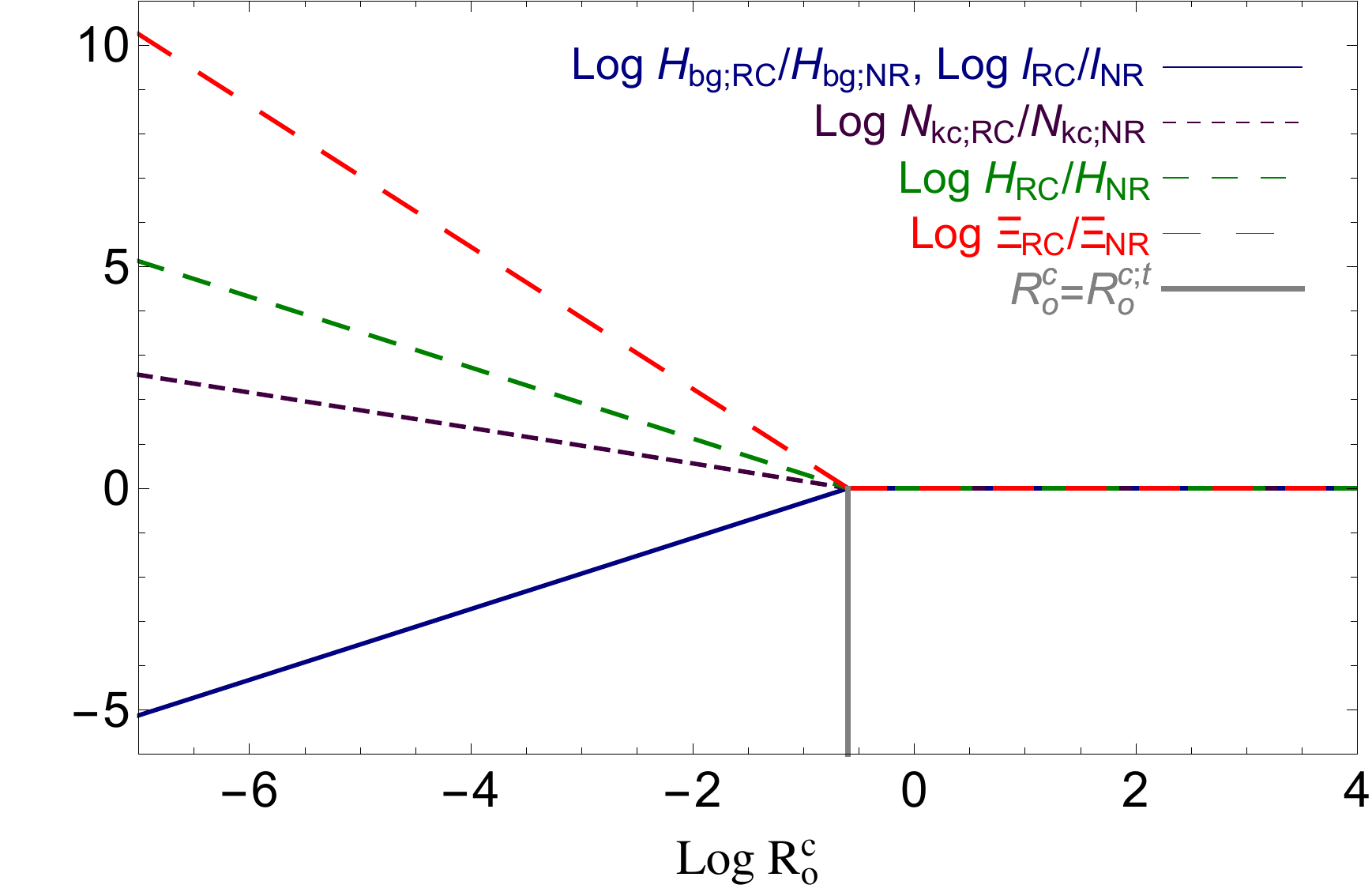}
\end{center}
\caption{{\bf Top:} Tidal dissipation frequency spectrum for the academic forcing chosen here \citep[see also][]{OL2004,ADMLP2015} for $E=10^{-4}$ and $\theta=0$ (in logarithm scale for the dissipation). {\bf Bottom:} {variations of the logarithm of the ratios $H_{\rm bg;RC}/H_{\rm bg;NR}$, $l_{\rm RC}/l_{\rm NR}$ (solid blue line), $N_{\rm kc;RC}/N_{\rm kc;NR}$ (dashed purple line), $H_{\rm RC}/H_{\rm NR}$ (dashed green line) and $\Xi_{\rm RC}/\Xi_{\rm NR}$ (long-dashed red line) as a function of $\log R_{\rm o}^{\rm c}$ when taking into account (or not) the action of rotation on turbulent friction.}}
\label{fig:spectra}
\end{figure}

Because of the form of the forced velocity field (Eq. \ref{v:sol}), the tidal dissipation spectrum ($\zeta$) is a complex resonant function of the normalized tidal frequency ($\omega$). It corresponds to resonances of the inertial waves that propagate in planetary and stellar convection zones. An example of such resonant spectra is computed in Fig. \ref{fig:spectra} (top panel) for $E=10^{-4}$ and $\theta=0$. We here use the academic forcing $f_{mn}=-i/(4\vert m\vert n^2)$, $g_{mn}=0$ and $h_{mn}=0$ adopted by \cite{OL2004} and \cite{ADMLP2015}. As discussed in \cite{ADMLP2015}, such an academic forcing already allows us to study properly the variation of $\zeta$ as a function of rotation, viscosity and frequency forcing. Following this latter work, we characterize $\zeta$ by the following physical quantities and scaling laws in the local model:
\begin{itemize}
\item the non-resonant background of the dissipation spectra $H_{\rm bg}$, which corresponds to the viscous dissipation of the equilibrium/non wave-like tide; it scales as $H_{\rm bg}\propto E$;
\item the number of resonant peaks $N_{\rm kc}$; it scales as $N_{\rm kc}\propto E^{-1/2}$;
\item their width at half-height $l_{mn}$; it scales as $l_{mn}\propto E$;
\item their height $H_{mn}$; it scales as $H_{mn}\propto E^{-1}$;
\item the sharpness of the spectrum defined as $\Xi=H_{11}/H_{\rm bg}$, which evaluate the contrast between the dissipation of the dynamical and equilibrium tides; it scales as $\Xi\propto E^{-2}$.
\end{itemize}
From now on, $X_{\rm RC}$ is a quantity evaluated with $E_{\rm RC}$ ({\it i.e.} with $\nu_{\rm T;RC}$) while $X_{\rm NR}$ is computed using $E_{\rm NR}$ ({\it i.e.} with $\nu_{\rm T;NR}$).\\

\subsection{The impact of rotation on the turbulent convective friction applied on tidal flows}
\label{subsec:tidalflows}

We can thus deduce interesting conclusions from obtained results in this simplified model for both the equilibrium and dynamical tides.
\subsubsection{The {\it equilibrium tide}} 
In our local Cartesian set-up, it is represented by the non-resonant background $H_{\rm bg}$. Using Eq. (\ref{def:ERC}), we thus deduce that its efficiency scales as $\Omega^{-9/5}$ in the regime of rapid rotation. This loss of efficiency of the equilibrium tide in rapidly rotating convective regions is illustrated in Fig. \ref{fig:spectra} where we plotted $\log\left(H_{\rm bg;RC}/H_{\rm bg;NR}\right)$ as a function of $\log R_{\rm o}^{\rm c}$ (in continuous blue line). 
\subsubsection{The {\it dynamical tide}} 
We use scaling laws obtained for the resonances of tidal inertial waves. We deduce that as soon as studied convective regions are in the regime of rapid rotation, their number and height respectively increase as $N_{\rm kc}\propto\Omega^{9/10}$ and $H_{mn}\propto\Omega^{9/5}$ while their width decreases as $l_{mn}\propto\Omega^{-9/5}$. The sharpness of $\zeta$ is increased as $\Xi\propto\Omega^{18/5}$. These variations of the properties of the resonant tidal dissipation frequency spectra is illustrated in Fig. \ref{fig:spectra} where we plot $\log\left(N_{\rm RC}/N_{\rm NR}\right)$, $\log\left(l_{\rm RC}/l_{\rm NR}\right)$, and $\log\left(\Xi_{\rm RC}/\Xi_{\rm NR}\right)$ as a function of $\log R_{\rm o}^{\rm c}$. 

As demonstrated by \cite{Aetal2014} \citep[see also][]{WS1999}, this may have important consequences for the evolution of the spin of the host body and of the orbits of the companions for example in the cases of star-planet and planet-moon systems. Indeed, the relative migration induced by a resonance scales as $\Delta a/a\equiv l_{mn}\,H_{mn}^{1/4}\propto \Omega^{-27/20}$. 

\section{Astrophysical discussion}

It is now important to discuss our results in the context of stellar and planetary interiors hosting turbulent convection regions. We focus here on the convective envelopes of low-mass stars (from M- to F-types) and of gaseous/icy giant planets and of the cores of telluric bodies.

\subsection{The case of low-mass and solar-type stars}

As demonstrated by \cite{Zahn1966}, \cite{OL2007} and \cite{Mathis2015}, external convective zones of low-mass stars are key contributors for the dissipation of tidal kinetic energy in these objects\footnote{\cite{Zahn1977} showed that tidal dissipation in intermediate-mass and massive stars is dominated by thermal diffusion acting on gravity waves in their radiative envelope.}. This is illustrated by the observations of the state of the orbits and of stellar spins in planetary systems and binary stars \citep[see the synthetic review in][and references therein]{Ogilvie2014}. An illustrative example is given by the need to understand the orbital properties of hot Jupiters and the inclination angle of their orbits relatively to the spin axis of their host stars \citep[e.g.][]{Winnetal2010,Albrechtetal2012,VR2014a,VR2014b}, which vary with their properties (i.e. their mass, their age, their rotation, etc.).

In this context, the rotation rate of low-mass stars strongly vary along their evolution \citep[e.g.][]{Barnes2003,GalletBouvier2013,GalletBouvier2015,Amardetal2016,McQuillanetal2014,Garciaetal2014}. They are first locked in corotation with their initial circumstellar accretion disk because of complex MHD star-disk-wind interactions \citep[e.g.][]{Mattetal2010,Mattetal2012,Ferreira2013}. Next, because of their contraction, they spin up along the Pre Main Sequence (PMS) until the Zero-Age Main Sequence (ZAMS). Finally, they spin down during the Main Sequence (MS) because of the torque applied by magnetized stellar winds \citep[e.g.][]{Schatzman1962,Kawaler1988,Chaboyeretal1995,Revilleetal2015,Mattetal2015}. These variations of the angular velocity strongly affect the properties of convective flows because of the action of the Coriolis acceleration. Indeed, rapid rotation during the PMS modifies and constrains convective turbulent flow patterns, large-scale meridional circulations and differential rotation \citep{Ballotetal2007,Brownetal2008}. The key control parameters to unravel such a complex dynamics is the convective Rossby number $R_{\rm o}^{\rm c}$ defined in Eq. (\ref{defRoc}). For example, it allows us to predict the latitudinal behavior of the differential rotation established by convection \citep{Mattetal2011,Gastineetal2014,Kapylaetal2014} while it controls the friction applied on tidal flows (see Eq. \ref{nuR}) and the properties of the frequency spectrum of theirs dissipation (see Sec. \ref{subsec:tidalflows}).

\begin{figure*}[ht!]
\begin{center}
\includegraphics[width=0.445\linewidth]{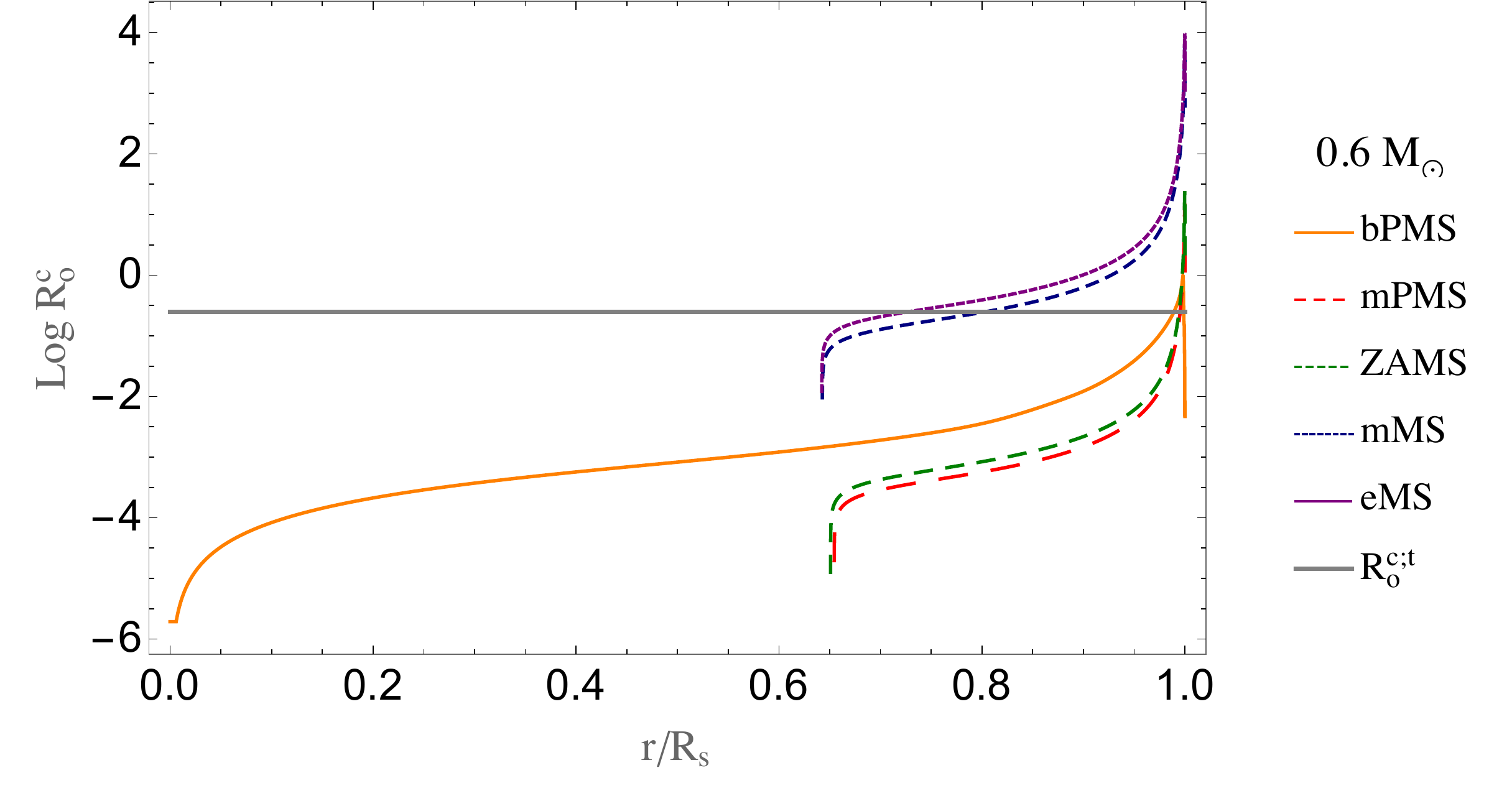}
\includegraphics[width=0.445\linewidth]{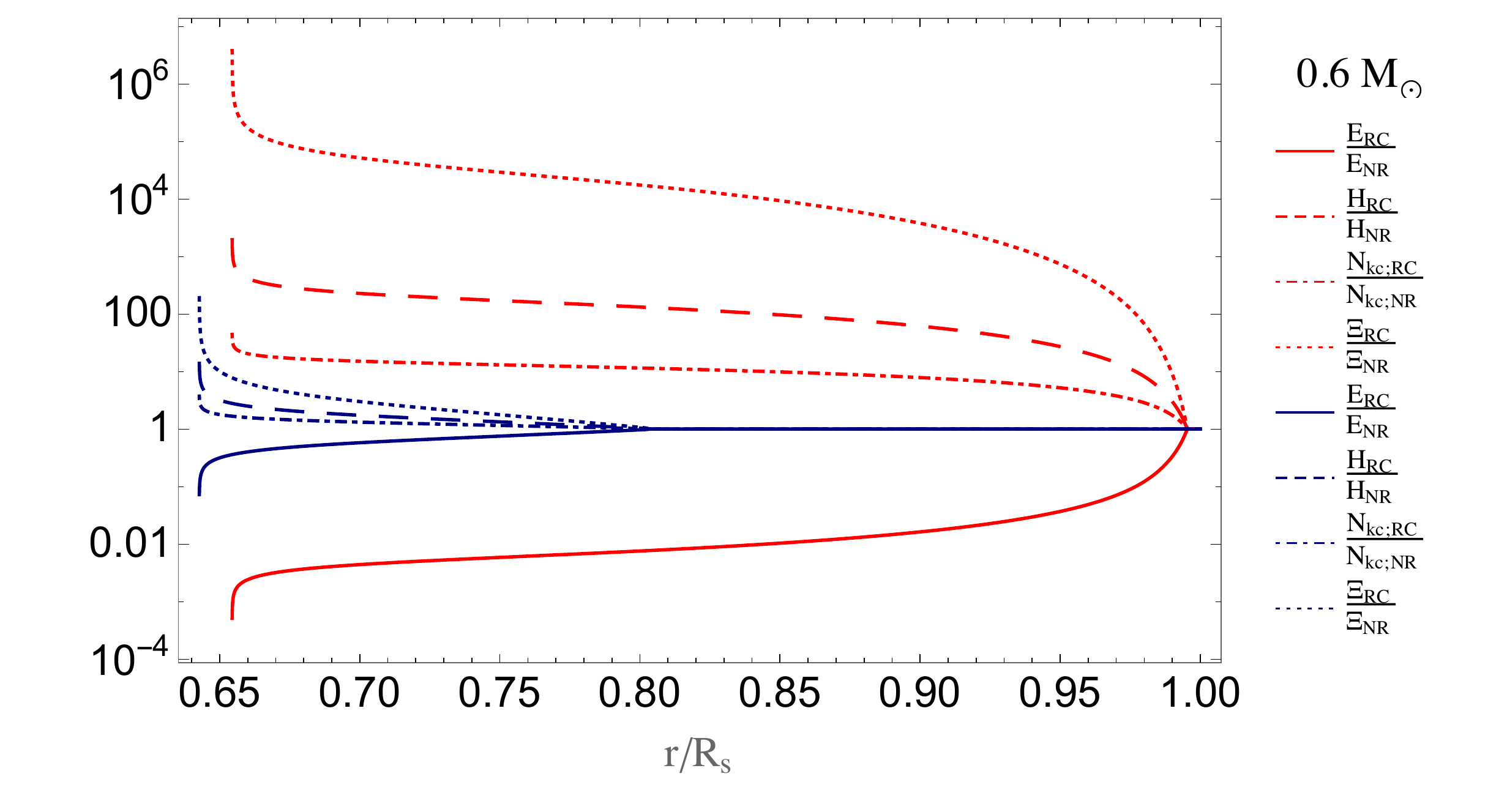}
\includegraphics[width=0.445\linewidth]{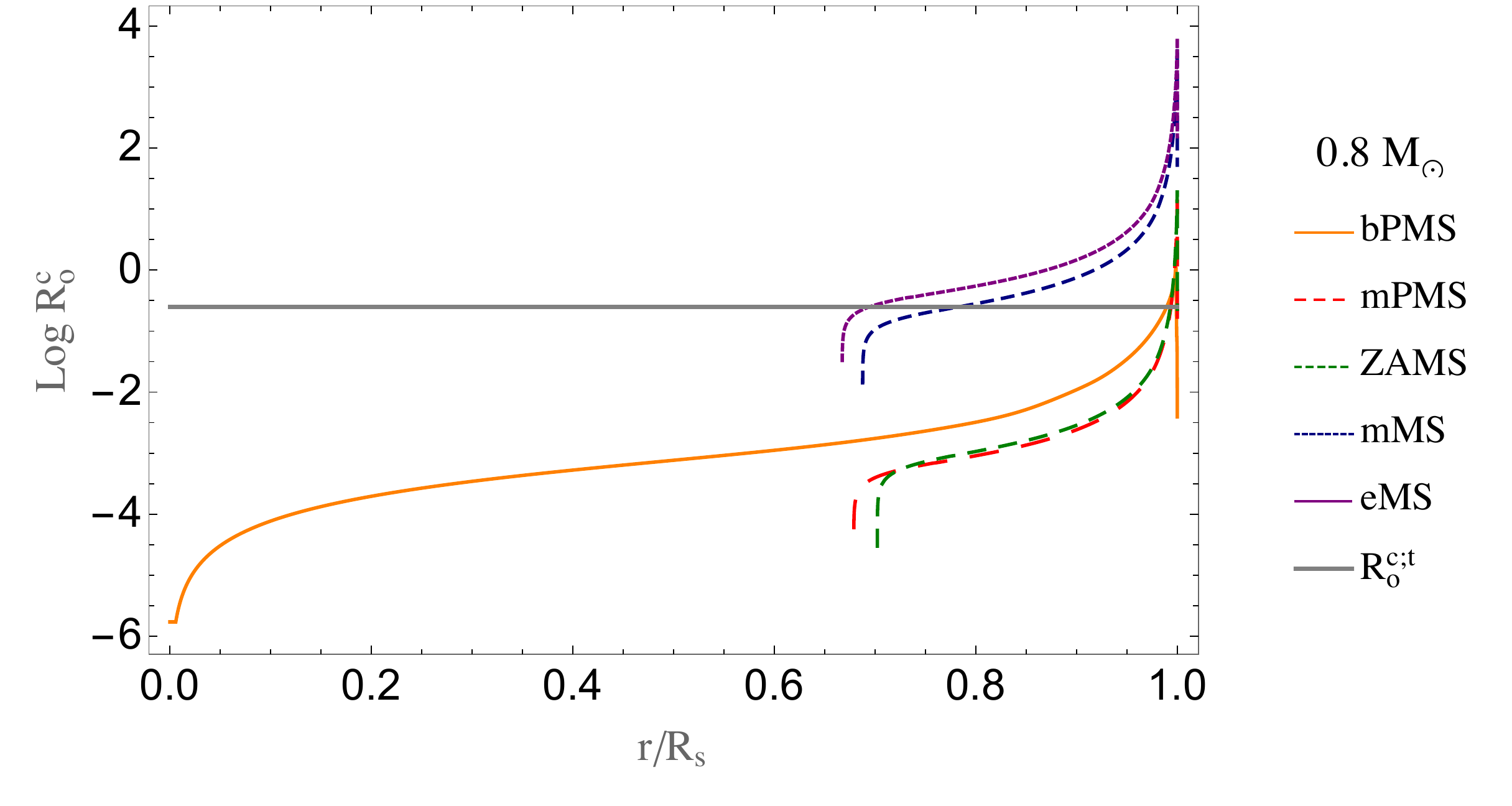}
\includegraphics[width=0.445\linewidth]{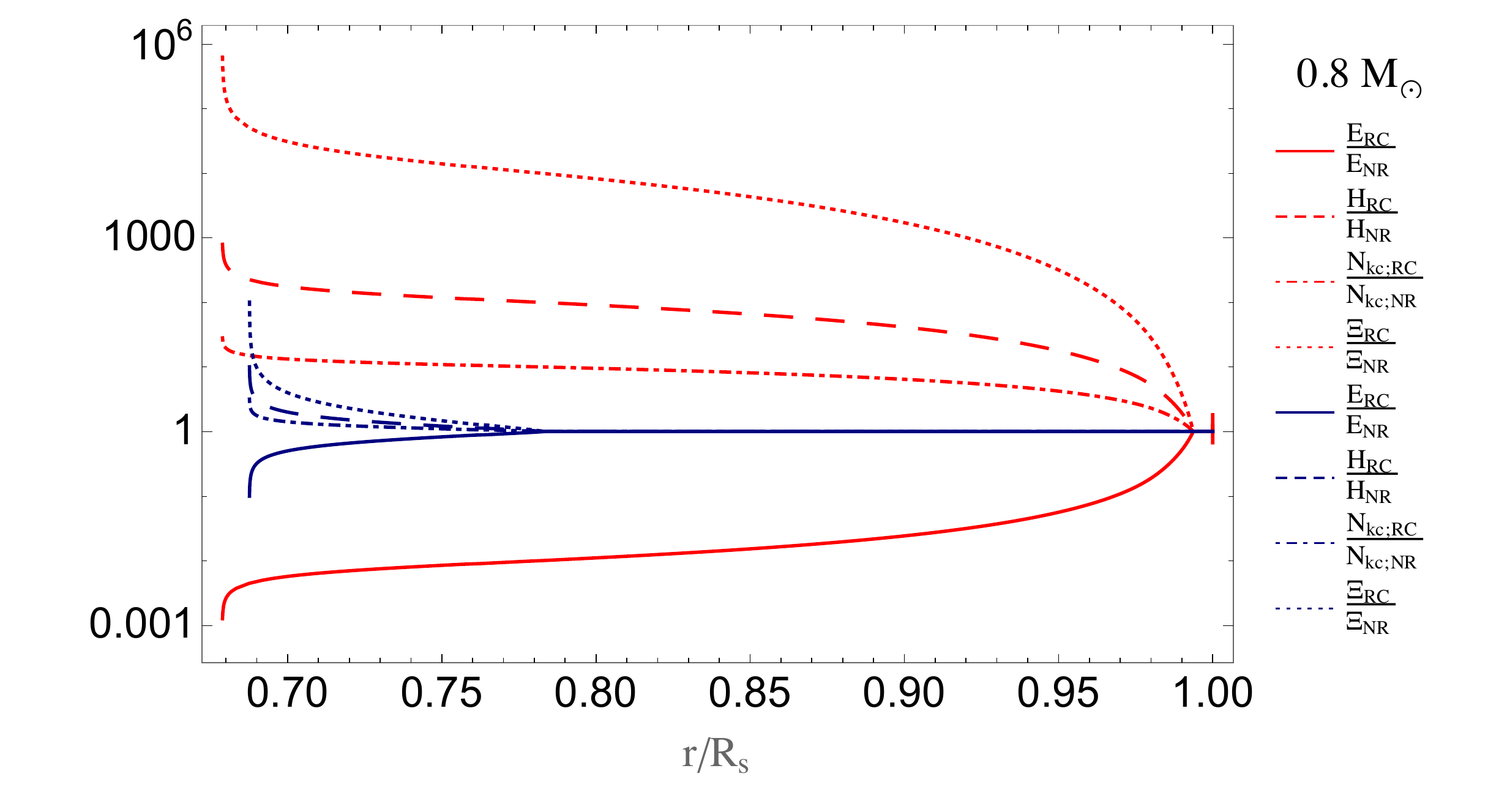}
\includegraphics[width=0.445\linewidth]{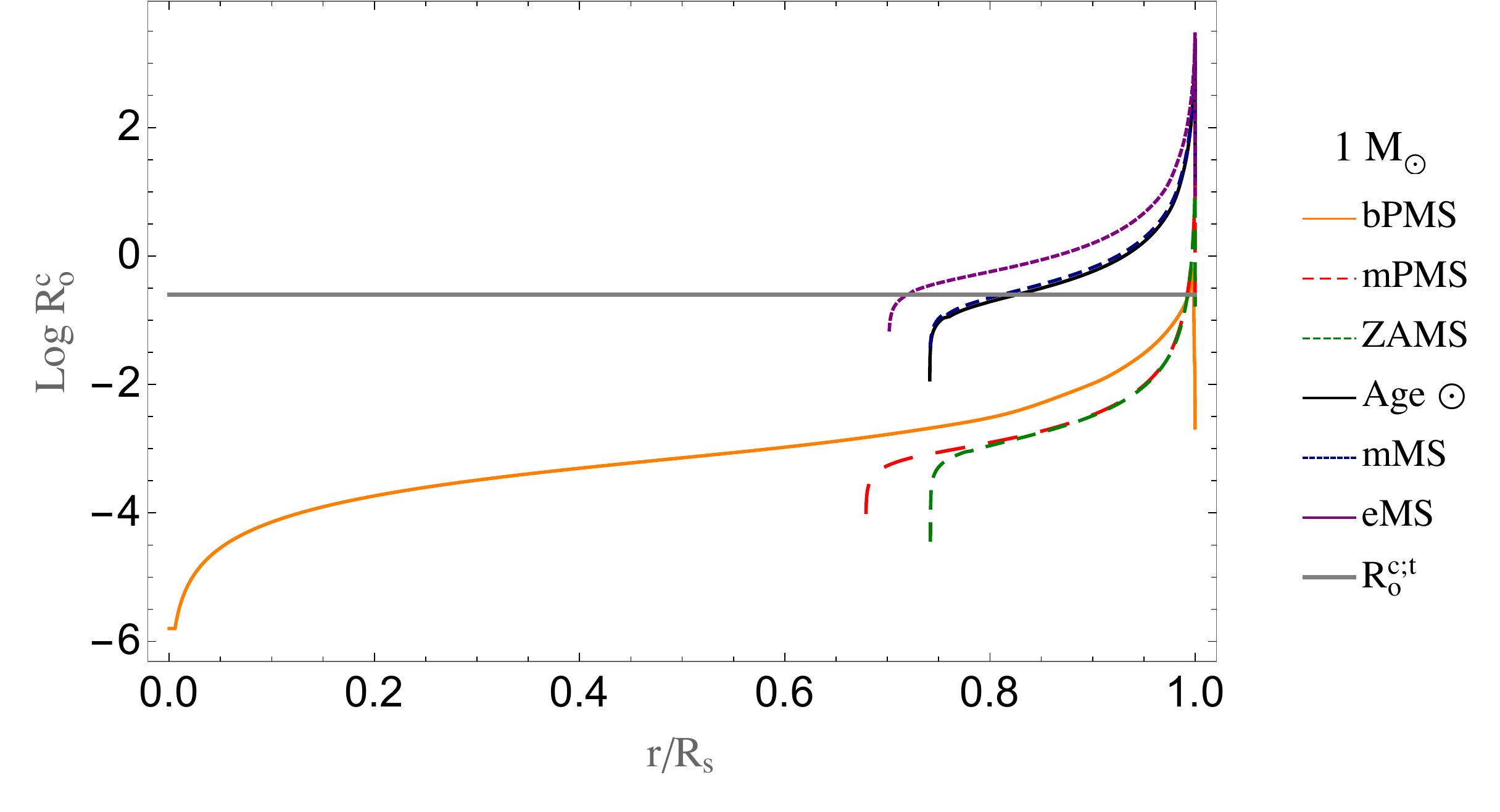}
\includegraphics[width=0.445\linewidth]{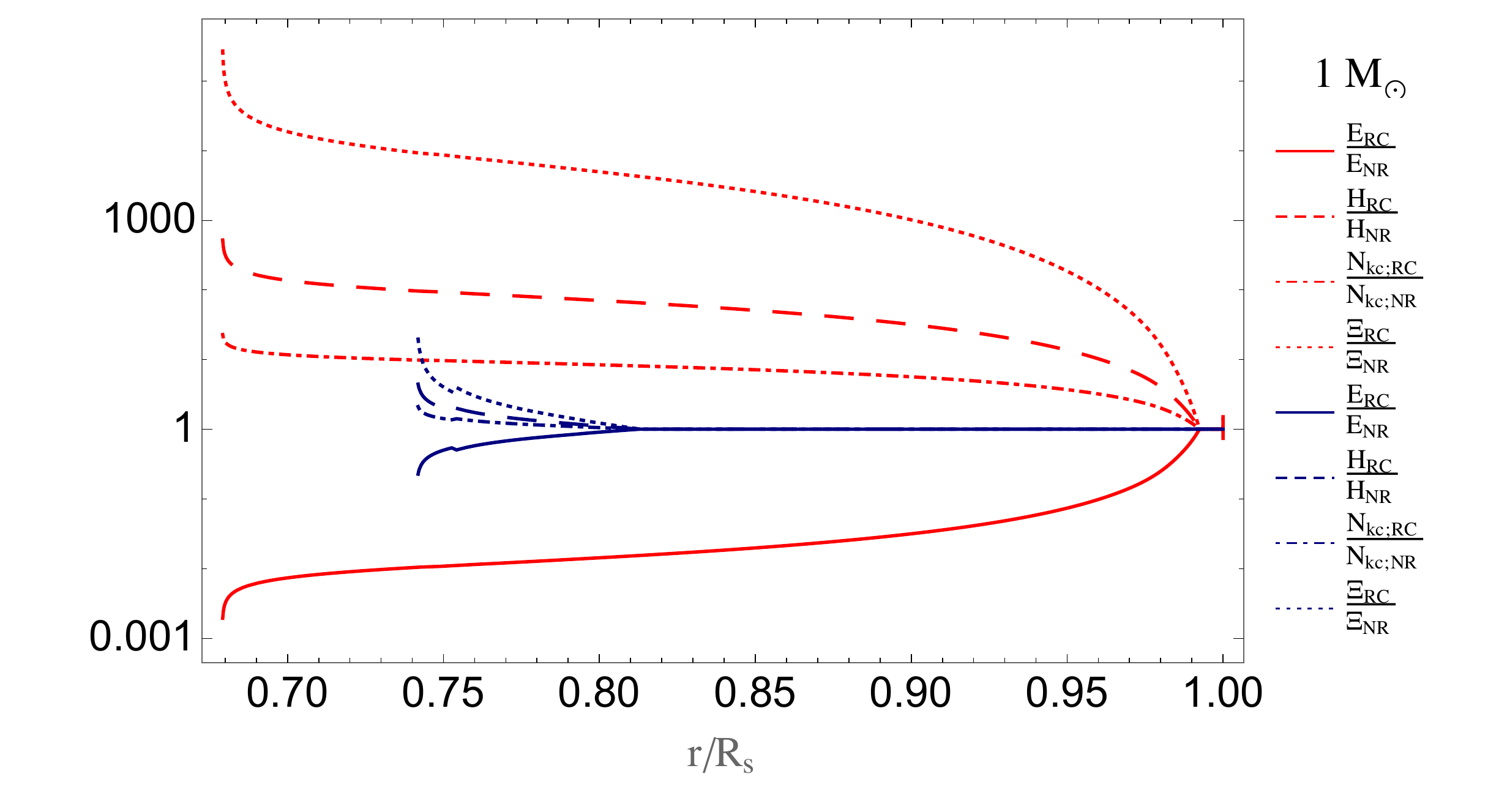}
\includegraphics[width=0.445\linewidth]{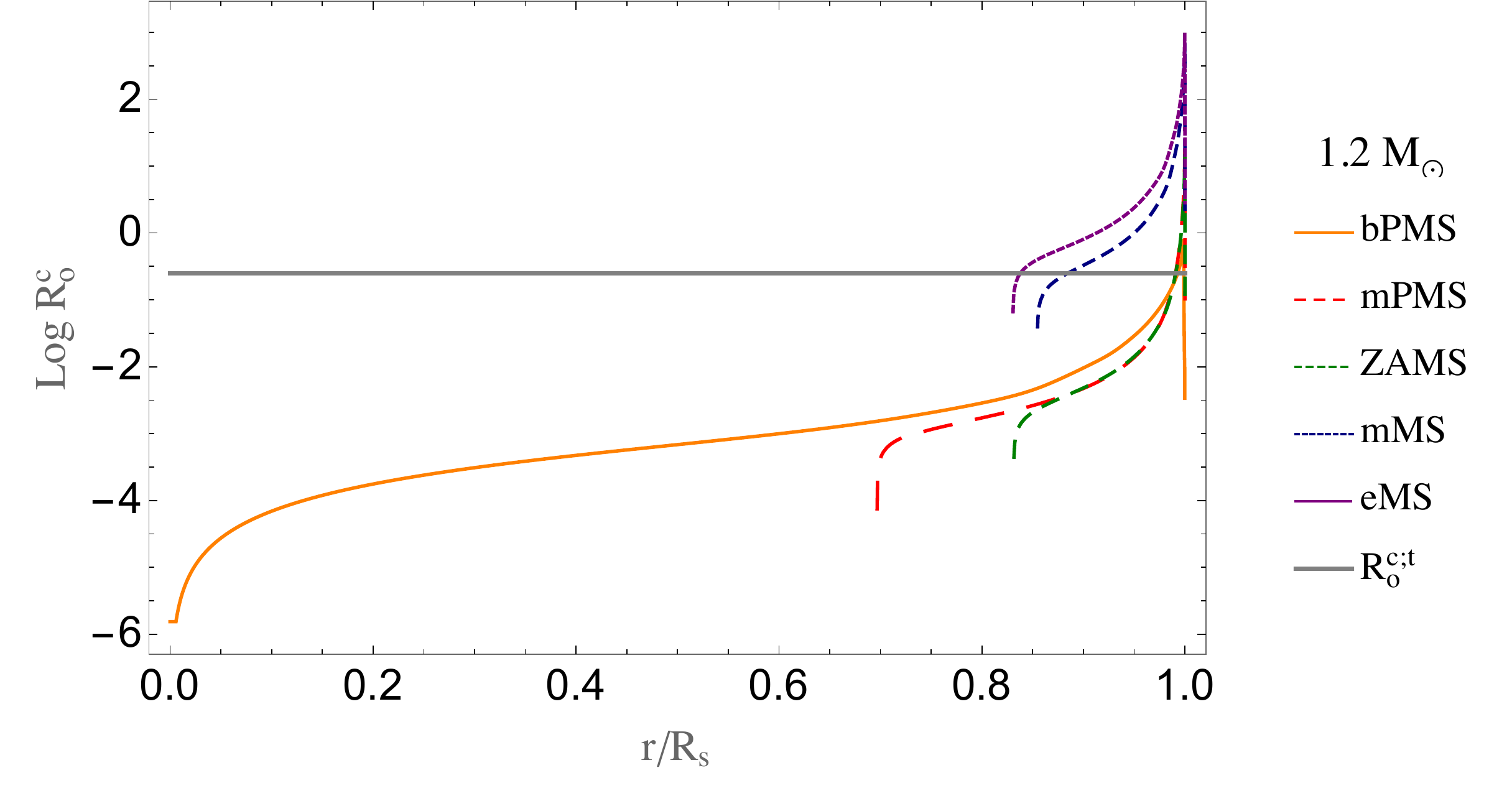}
\includegraphics[width=0.445\linewidth]{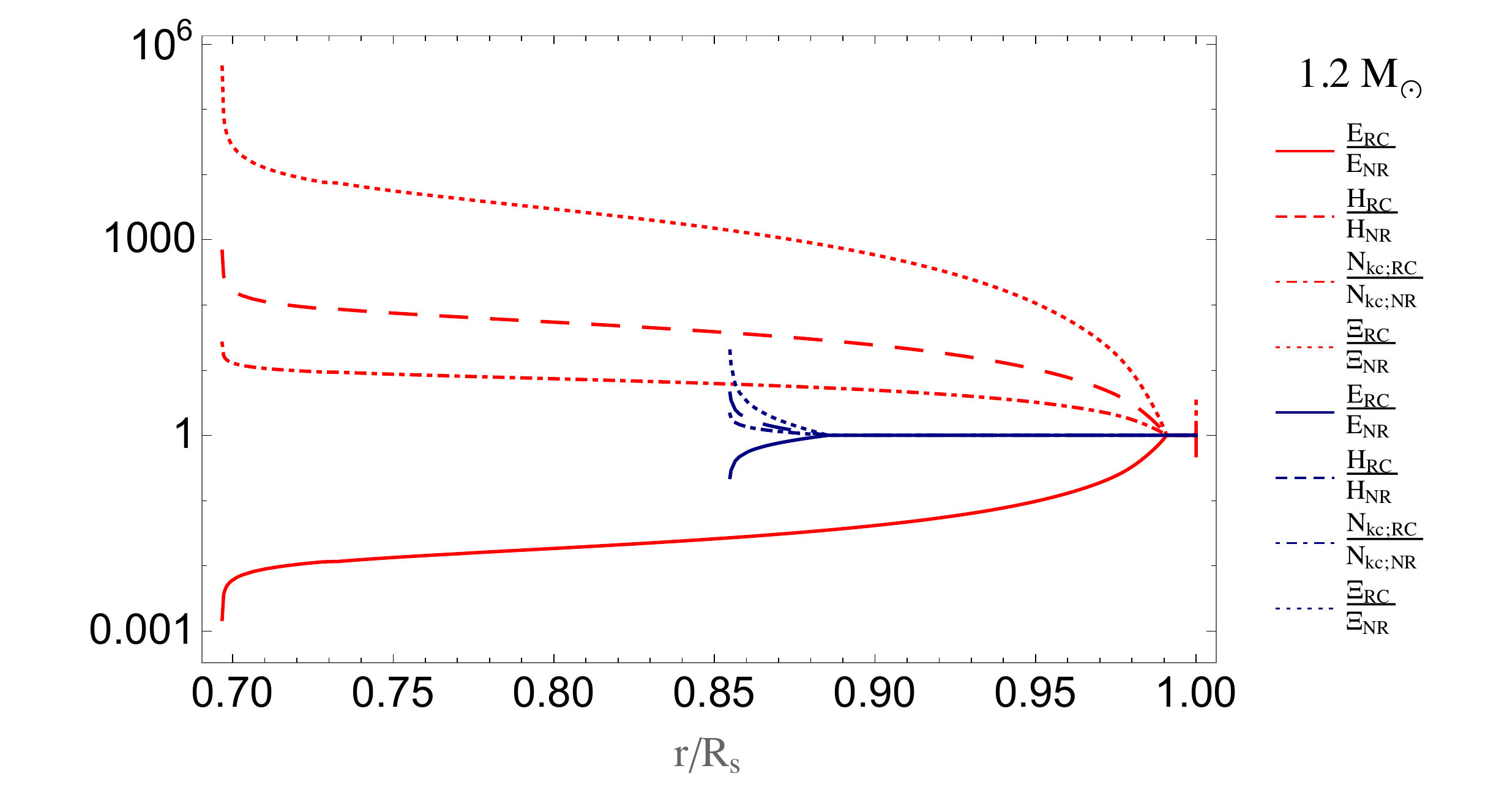}
\end{center}
\caption{{\bf Left panels:} radial profiles of $\log R_{\rm o}^{c}$ for $\left\{0.6,0.8,1,1.2\right\}M_{\odot}$ stars at different ages: the beginning of the PMS (in solid orange line; noted bPMS), the mid PMS (in dashed red line; noted mPMS), the ZAMS (in dashed green line), the middle of the MS (in dashed blue line; noted mMS), the end of the MS (in dashed purple line; noted eMS) and the solar age for the $1M_{\odot}$ solar-type star (in solid black line; noted Age $\odot$); $r/R_{\rm s}$ is the normalized radius, where $R_{\rm s}$ is the radius of the star. The critical convective Rossby number $R_{\rm o}^{\rm c;t}\approx0.25$ giving the transition between the rapidly and the slowly rotating regimes is represented by the solid thick grey line. {\bf Right panels:} radial profiles of $E_{\rm RC}/E_{\rm NR}$ (and $\nu_{\rm T;RC}/\nu_{\rm T;NR}$, $H_{\rm bg;RC}/H_{\rm bg;NR}$ and $l_{\rm RC}/l_{\rm NR}$; in solid line), $H_{\rm bg;RC}/H_{\rm bg;NR}$ (in dashed line), $N_{\rm RC}/N_{\rm NR}$ (in dot-dashed line) and $\Xi_{\rm RC}/\Xi_{\rm NR}$ (in dotted line) for each stellar mass at the mid PMS (in red) and mid MS (in blue).}
\label{RoProfiles}
\end{figure*}

\begin{figure}[h!]
\begin{center}
\includegraphics[width=\linewidth]{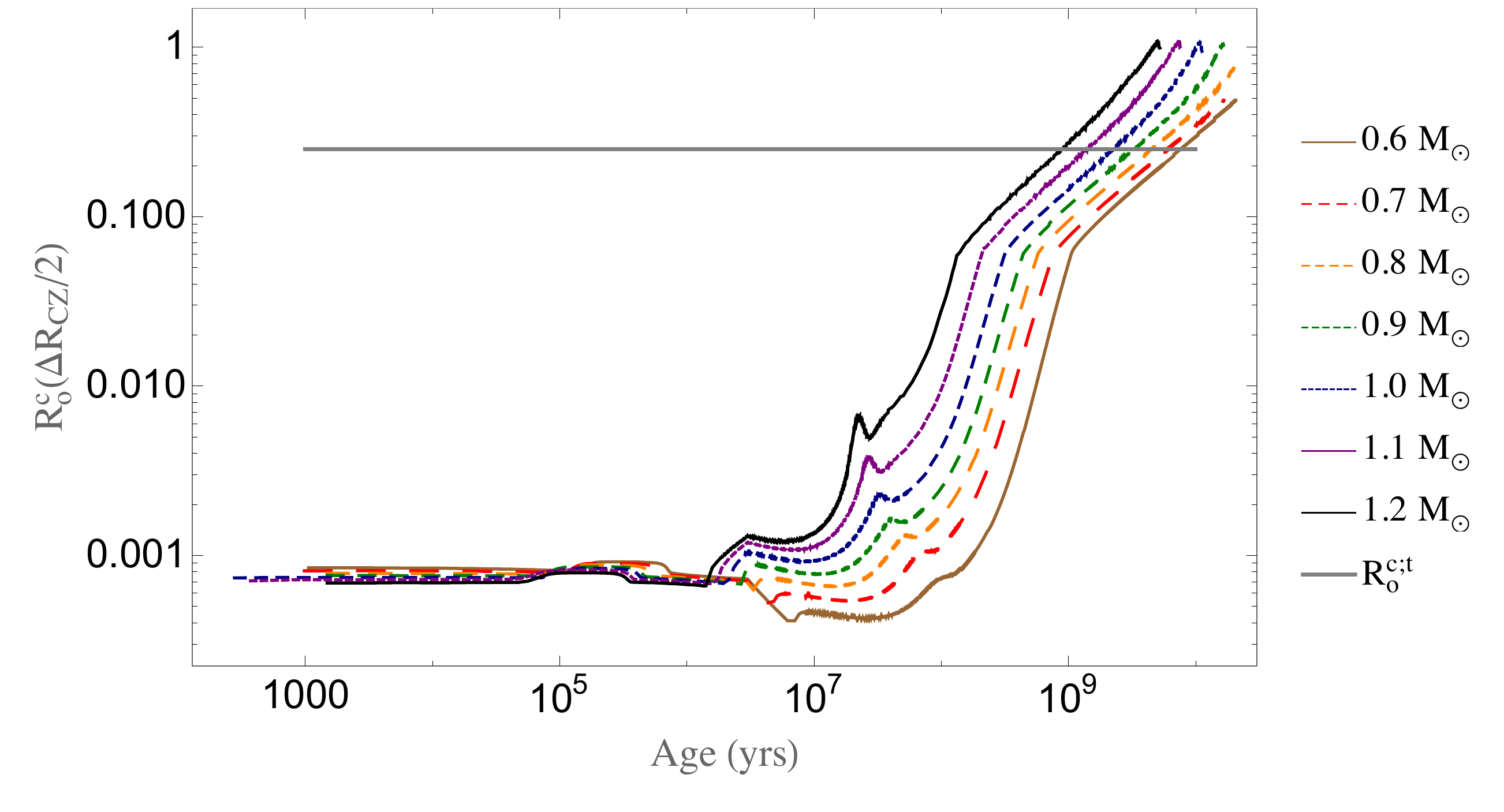}
\caption{{Evolution of $R_{\rm o}^{\rm c}$, evaluated in the middle of the convective envelope (at $r=\Delta R_{\rm CZ}/2$), as a function of stellar age (in logarithm scales) computed using the grid of rotating stellar models} 
for 0.6M$_{\odot}$ (dashed brown line), 0.7M$_{\odot}$ (dashed red line), 0.8M$_{\odot}$ (dashed orange line), 0.9M$_{\odot}$ (dashed green line), 1M$_{\odot}$ (dashed blue line), 1.1M$_{\odot}$ (dashed purple line) and 1.2M$_{\odot}$ (solid black line) stars. The critical convective Rossby number $R_{\rm o}^{\rm c;t}\approx0.25$ giving the transition between the rapidly and the slowly rotating regimes is represented by the solid thick grey line.}
\label{RoES}
\end{center}
\end{figure}

\begin{figure*}[t!]
\begin{center}
\includegraphics[width=0.4975\linewidth]{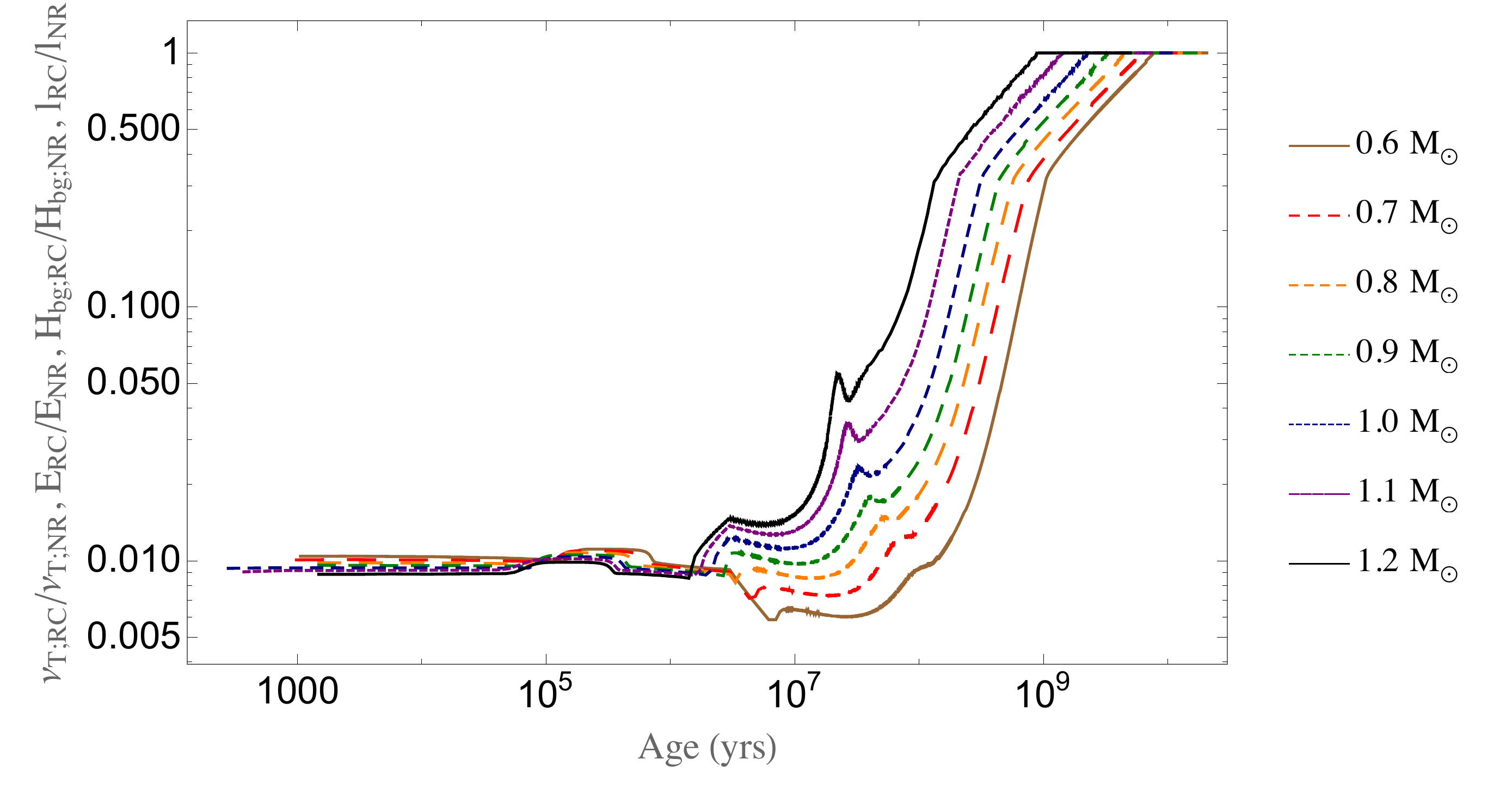}
\includegraphics[width=0.4975\linewidth]{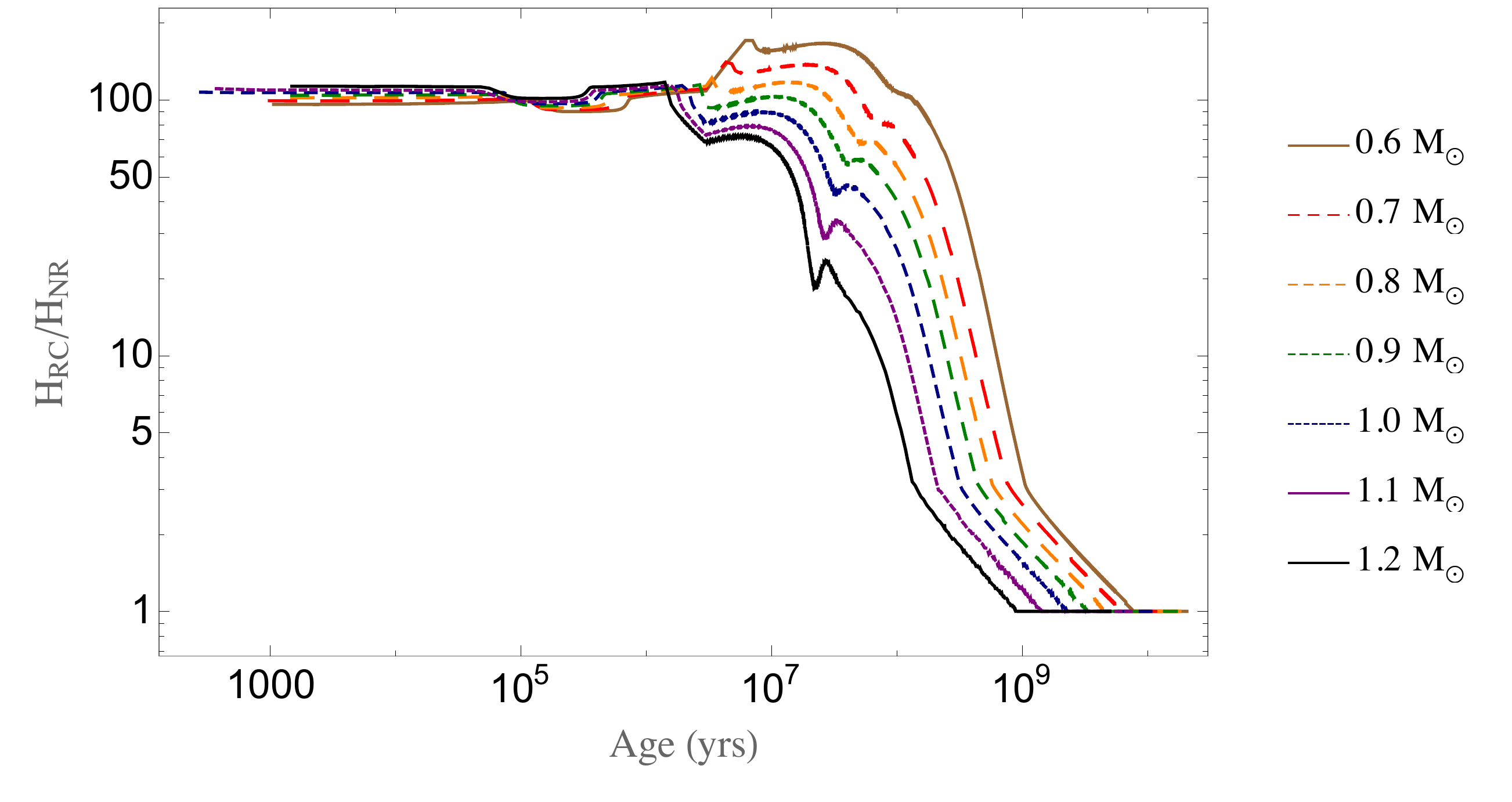}
\includegraphics[width=0.4975\linewidth]{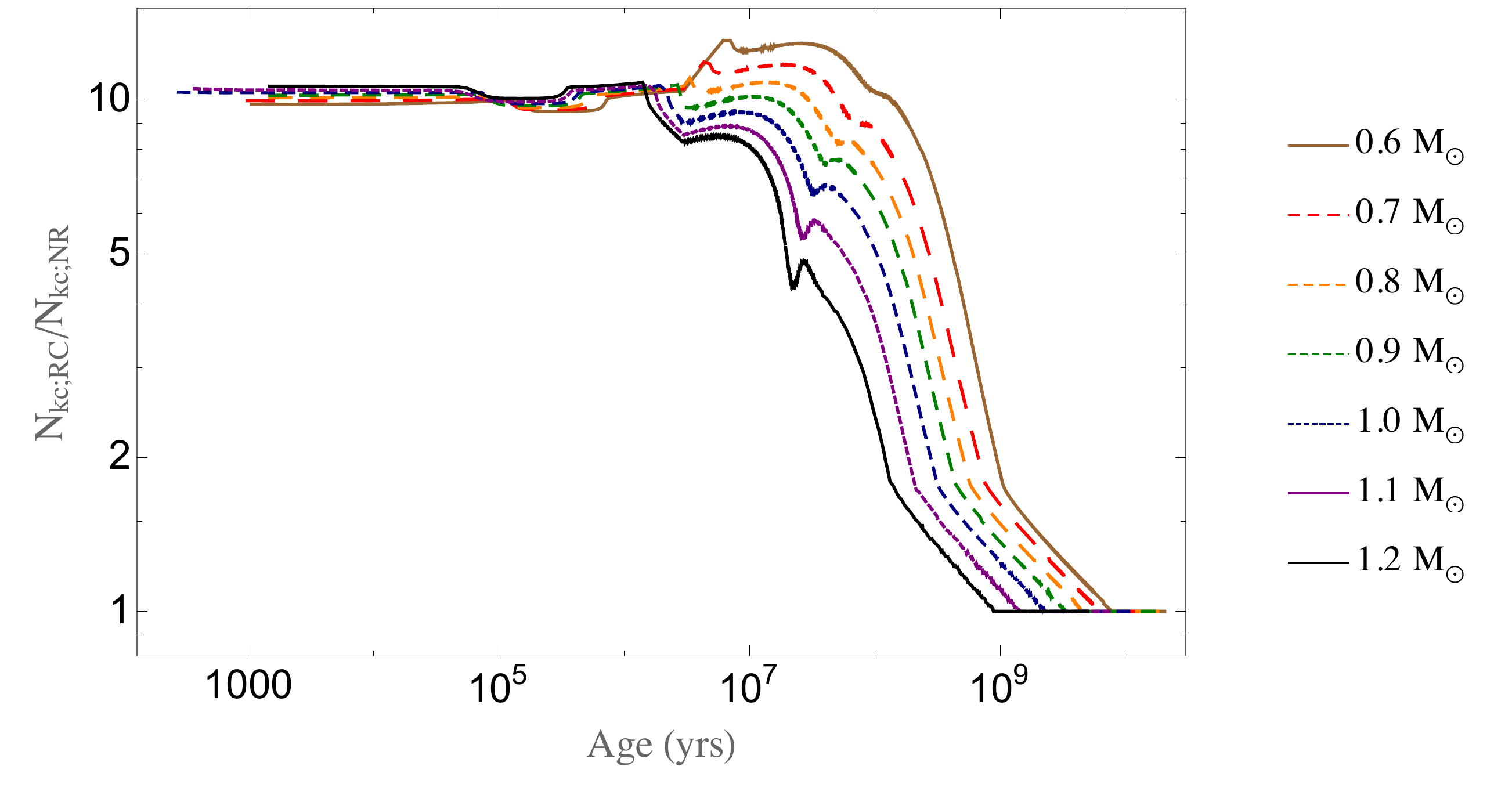}
\includegraphics[width=0.4975\linewidth]{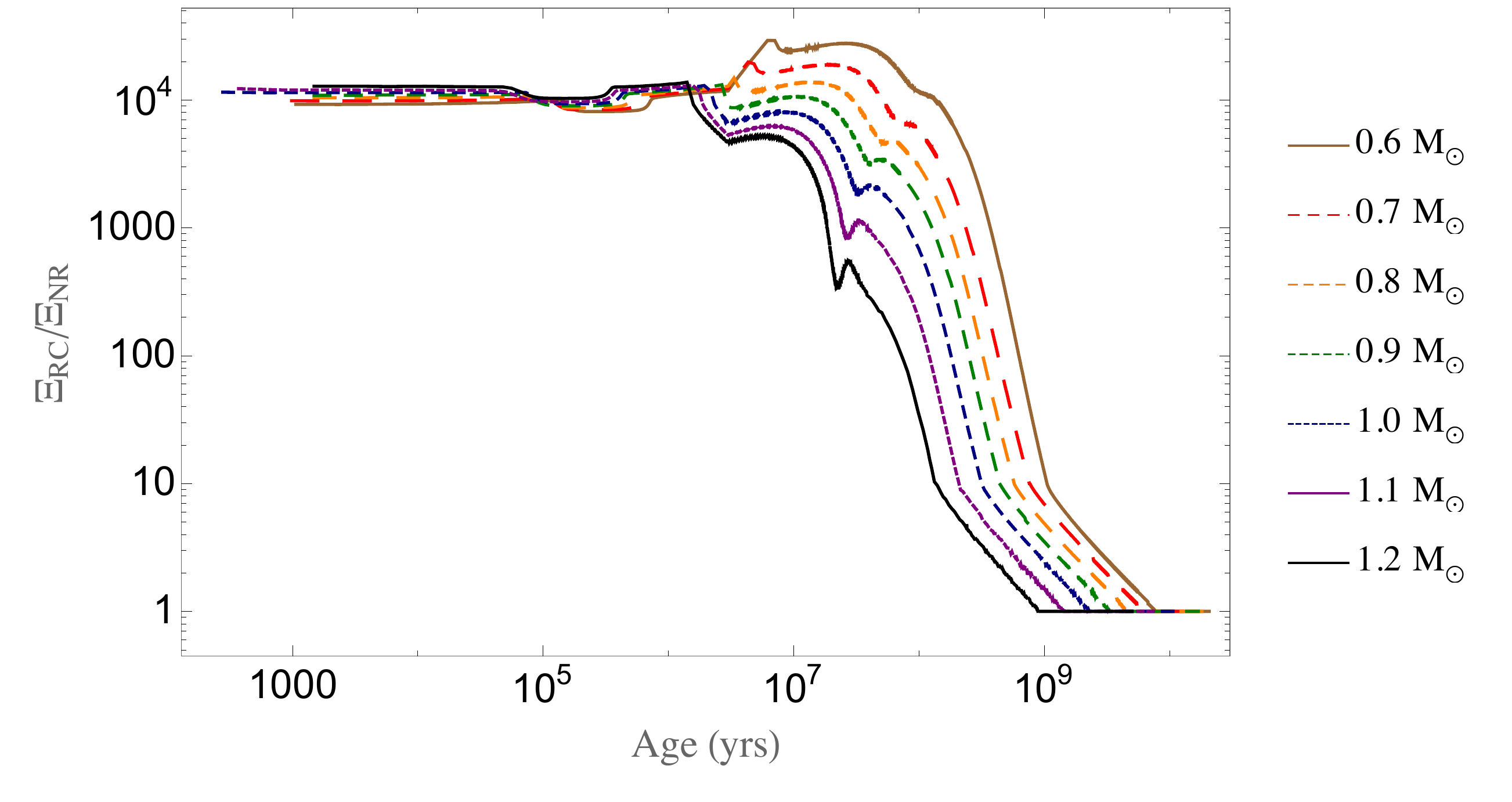}
\caption{{\bf Top left:} Evolution of $\nu_{\rm T;RC}/\nu_{\rm T;NR}$, $E_{\rm RC}/E_{\rm NR}$, $H_{\rm bg;RC}/H_{\rm bg;NR}$ and $l_{\rm RC}/l_{\rm NR}$ at $r=\Delta R_{\rm CZ}/2$ (not written on labels to lighten notations) as a function of stellar age (in logarithm scales) for stellar masses going from 0.6M$_{\odot}$ to 1.2M$_{\odot}$ (colors have been in defined in caption of Fig. \ref{RoES}). {\bf Top right:} same for $H_{\rm RC}/H_{\rm NR}$. {\bf Bottom left:} same for $N_{\rm RC}/N_{\rm NR}$. {\bf Bottom right:} same for $\Xi_{\rm RC}/\Xi_{\rm NR}$.}
\label{SLES}
\end{center}
\end{figure*}

It is thus interesting to compute the variations of $R_{\rm o}^{\rm c}$ along the evolution of low-mass stars. It will allow us to evaluate the impact of the action of rotation on the turbulent convective friction and on the properties of the tidal dissipation along stellar evolution thanks to Eqs. (\ref{nuR}-\ref{RR}) and to scaling laws given in Sec. \ref{sec:scalinglaws}. To reach this objective, we choose to {compute new grids of stellar rotating models for stars with masses between $0.6{\rm M}_{\odot}$ and $1.2{\rm M}_{\odot}$ (i.e. from K- to F-types) with a solar metallicity. We follow the methodology first proposed by \cite{Landinetal2010}. This allows us to study the variations of the rotation and $R_{\rm o}^{\rm c}$ as a function of time and radius along the evolution of these stars from the beginning of the PMS to the end of the MS. We use the latest version of the STAREVOL stellar evolution code described in details (e.g. for the used equation of state, nuclear reaction, opacities, etc.) in \cite{Amardetal2016} \citep[see also][]{Siessetal2000,Palaciosetal2003,Lagardeetal2012}. The mixing-length parameter is chosen to be $\alpha=1.6267$. Convective regions are modeled assuming uniform rotation while radiation zones rotate according to redistribution of angular momentum by shear-induced turbulence and meridional flows, which are treated using formalisms derived by \cite{Zahn1992,MaederZahn1998,MathisZahn2004,MPZ2004}.} Angular momentum losses at the stellar surface due to pressure-driven magnetized stellar winds are taken into account using the prescription adopted by \cite{Mattetal2015} {(the two parameters of this model are given in Eqs. 6 \& 7 of this paper, and we choose $\chi=10$ and $p=2$)}, which allows the authors to verify the Skumanich law \citep{Skumanich1972}. {Since our interest here is on the impact of rapid rotation, we choose to compute the evolution of initially rapidly rotating low-mass stars, which all have a same initial rotation period of $1.4$ days and disc-locking time of $3\times10^6$ years. In addition, we also have computed the evolution of initially slow and median rotators that allows us to compare and validate our results with those obtained by \cite{Landinetal2010}}. 

{In Fig. \ref{RoProfiles} (left panel), we first represent the radial profiles of $\log R_{\rm o}^{\rm c}$ for different masses (i.e. $0.6,0.8,1.0$ and $1.2M_{\odot}$) and ages: the beginning of the PMS (in solid orange line), the mid PMS (in dashed red line), the ZAMS (in dashed green line), the middle of the MS (in dashed blue line), the end of the MS (in dashed purple line) and the solar age for the $1M_{\odot}$ solar-type star (in solid black line) \footnote{Note here that we consider a $1M_{\odot}$ solar-type star and not a helioseismic calibrated solar model.}. In these plots (and in Fig. \ref{RoES}), we also recall the value of the critical convective Rossby number $R_{\rm o}^{\rm c;t}\approx0.25$. It corresponds to the transition between the rapidly ($R_{\rm o}^{\rm c}<R_{\rm o}^{\rm c;t}$) and slowly ($R_{\rm o}^{\rm c}>R_{\rm o}^{\rm c;t}$) rotating regimes and is reported here by the solid thick grey line. For all stellar masses and ages, a large radial variation of $R_{\rm o}^{\rm c}$ over several ordres of magnitude is obtained with a monotonic increase towards the surface. This result is coherent with previously published results in the case of the Sun \citep[see Fig. 1 in][]{Kapylaetal2005}. Therefore, the impact of rotation on the turbulent friction applied on tidal flows may be stronger close to the basis of the convective envelope than in surface regions. This may impact differently equilibrium and dynamical tides. On one hand, the equilibrium tide is varying at the zeroth-ordre as $r^2Y_{2}^{M}\left(\theta,\varphi\right)$, where $Y_{2}^{M}$ is the quadrupolar spherical harmonics \citep[see e.g.][]{Zahn1966,RMZ2012}. On the other hand, the dynamical tide constituted by tidal inertial waves propagate in the whole convective region. In the case of fully convective low-mass stars (at the beginning of the PMS or in M-type stars), inertial waves propagate as regular modes \citep{Wu2005}. In the case of convective shells, they propagate along waves' attractors \citep{OL2007}. Therefore, for a given age, the turbulent friction applied on tidal inertial waves propagating deep inside convective zones would be more strongly affected by the action of the Coriolis acceleration on convective flows than the one applied on the equilibrium tide. However, for all stellar masses, we can see that the impact of rotation on the friction stays important everywhere during all the PMS and the beginning of the MS where stars are rotating rapidly. Indeed, during these evolution phases, the plotted radial profiles show that $R_{\rm o}^{\rm c}<R_{\rm o}^{\rm c;t}$ for all radii except just below the surface layers. This is confirmed in Fig. \ref{RoES}, where the convective Rossby number in the middle of the convective envelope (i.e. at $r=\Delta R_{\rm CZ}/2=(R_{\rm s}-R_{\rm c})/2$, where $R_{\rm s}$ and $R_{\rm c}$ are the radius of the star and those of the basis of the convective envelope respectively) is plotted as a function of time. 
Indeed, Fig. \ref{RoProfiles} (left panel) shows how $R_{\rm o}^{\rm c}\left(\Delta R_{\rm CZ}/2\right)$ can provide a reasonable intermediate ordre of magnitude for $R_{\rm o}^{\rm c}$ in tidal dissipation studies.} At these evolutionary stages, we thus expect for a strong action of the dynamical tide, constituted by inertial waves, with associated highy resonant tidal dissipation frequency-spectra. This result is very important since \cite{ZahnBouchet1989} demonstrated that the most important phase of orbital circularization in late-type binaries occurs during the PMS where the convective envelopes of the components are thick. Moreover, it perfectly matches the results obtained by \cite{Mathis2015} and \cite{BolmontMathis2016} where a strong action of tidal inertial waves all along the PMS has been identified. In Figs. \ref{RoProfiles} (right panels) \& \ref{SLES}, we represent the ratios between the values of the turbulent eddy-viscosity (and of the corresponding Ekman number) and those of the dissipation frequency spectrum properties ($H_{\rm bg}$, $l$, $H$, $N_{\rm k_c}$ and $\Xi$) {(respectively as a function of $r$ for $\left\{0.6,0.8,1,1.2\right\}M_{\odot}$ stars at the middle of the PMS and MS in Fig. \ref{RoProfiles} (right panels) and as a function of time for $r=\Delta R_{\rm CZ}/2$ for stellar masses from $0.6$ to $1.2M_{\odot}$ in Fig. \ref{SLES}}) when taking into account the modification of the turbulent friction by rotation or not. In each case, we see that differences by several ordres of magnitude can be obtained for each quantity that must be taken into account. {It demonstrates that} the rapidly-rotating regime enhances the highly resonant dynamical tide while it decreases the efficiency of the equilibrium tide because the amplitudes of their viscous dissipation increases and decreases respectively with $\nu_{\rm T}$. Therefore, it strengthens the {general} conclusion that tidal evolution of star-planet systems and of binary stars must be treated taking into account the dynamical tide and not only the equilibrium tide \citep[e.g.][]{Aetal2014,WS2002a,WS2002b}. As a consequence, the predictions obtained using simplified equilibrium tide models such as the constant tidal quality factor and the constant tidal lag models \citep[e.g.][]{Kaula1964,Hut1981} must be considered very carefully. As a conclusion, these results demonstrate how it is necessary to have an integrated and coupled treatment of the rotational evolution of stars, which directly impacts tidal dissipation in their interiors, and of the tidal evolution of the surrounding planetary or stellar systems \citep[e.g.][]{Penevetal2014,BolmontMathis2016}. In addition, tidal torques may also modify the rotation of stars as a back reaction \citep[e.g.][]{Bolmontetal2012,BolmontMathis2016}.

\begin{figure*}[t!]
\begin{center}
\includegraphics[width=0.4975\linewidth]{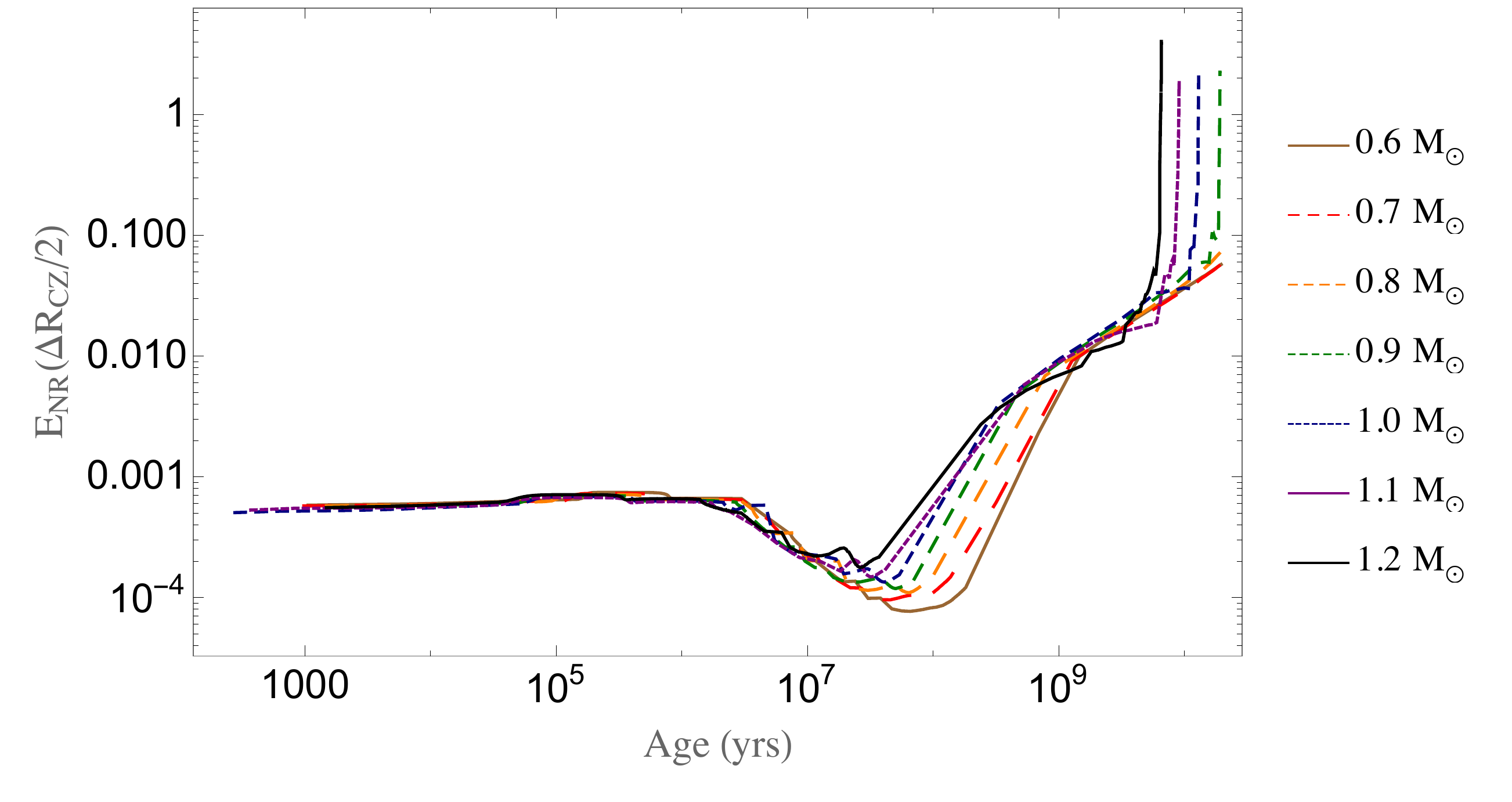}
\includegraphics[width=0.4975\linewidth]{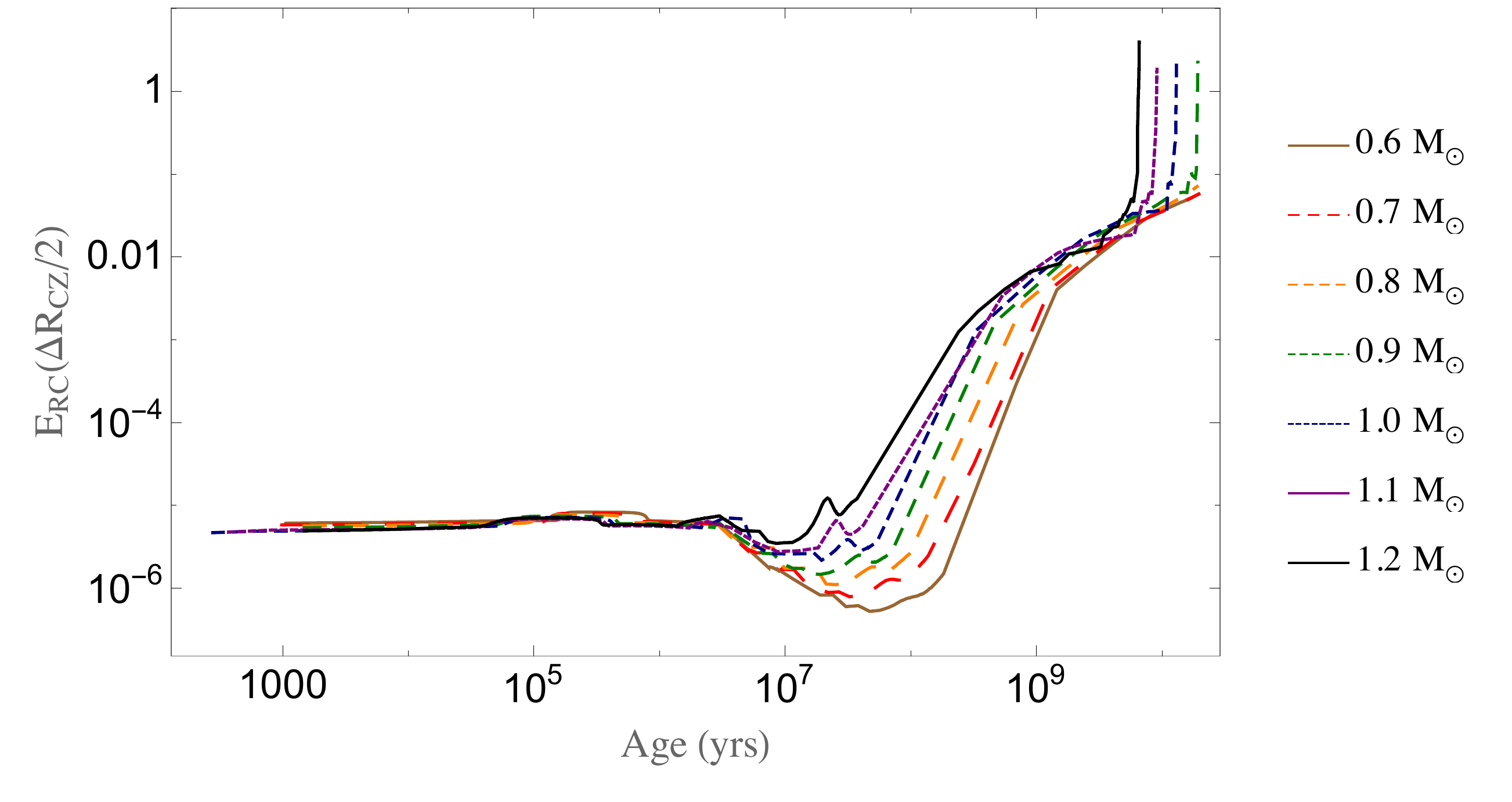}
\caption{Evolution of the Ekman numbers $E_{\rm NR}$ (left panel) and $E_{\rm RC}$ (right panel) at $r=\Delta R_{\rm CZ}/2$ as a function of stellar age (in logarithm scales) for stellar masses going from 0.6M$_{\odot}$ to 1.2M$_{\odot}$ (colors have been in defined in caption of Fig. \ref{RoES}).} 
\label{EkAge}
\end{center}
\end{figure*}

{Finally, variations of $E_{\rm NR}$ and $E_{\rm RC}$ (with $L=R_{\rm s}$; see Eqs. \ref{def:ENR} and \ref{def:ERC}) evaluated at $r=\Delta R_{\rm CZ}/2$ as a function of stellar mass and age are given in Fig. \ref{EkAge} with $f$ (in Eq. \ref{nuR}) fixed to $1$, in left and right panels respectively. STAREVOL models allow us to compute coherently $V_{\rm c}$, $L_{\rm c}$ and $R_{\rm o}^{\rm c}$ (Eq. \ref{nuR}) for each stellar mass, age and radius in the star. In these plots, we identify the phase of corotation of the stars with the surrounding disk (until $3\times10^6$ years), the acceleration phase, and the stellar spin-down due to the breaking by the wind. This provides key inputs for future numerical simulations of tidal dissipation in spherical convective shells for which the Ekman number is one of the key physical control parameter \citep[e.g.][]{OL2007,BR2013,GBMR2016}. It allows us to improve step by step the evaluation of tidal friction in the convective envelope of low-mass stars all along their evolution by combining hydrodynamical studies and stellar modeling.}



\subsection{The case of planets}

As in the case of young low-mass stars, convective flows are strongly constrained by the Coriolis acceleration in rapidly rotating planetary interiors. The most evident cases are those of the Earth liquid core and of the envelopes of gaseous (Jupiter and Saturn) and icy (Uranus and Neptune) giant planets \citep{Glatzmaier2013}. In this context, the importance of rotation on tidal flows and their dissipation has been demonstrated both in the case of the Earth \citep[e.g.][]{Buffett2010} and of giant planets \citep[e.g.][]{OL2004,Wu2005}. 

Therefore, as in the case of stars, it is interesting to collect values of the convective Rossby numbers for planetary interiors. We choose here to use the ordres of magnitude given in \cite{SchubertSoderlund2011} for the Earth and in \cite{Soderlundetal2013} for giant planets. They are reported in the following table.

\begin{table}[h!]
  \centering
  \begin{tabular}[center]{cccccc}\hline \hline
    Parameter & Earth & Jupiter & Saturn & Uranus & Neptune \\ \hline
    $R_{\rm o}^{\rm c}$ & $10^{-7}$ & $10^{-9}$  & $10^{-8}$  & $10^{-5}$  & $10^{-5}$ \\
    \hline\\
\end{tabular}
  \caption{Ordres of magnitude for planetary Rossby numbers (given in \cite{SchubertSoderlund2011} for the Earth and in \cite{Soderlundetal2013} for giant planets).}
  \label{tab:RoPlanet}
\end{table}

From our previous results, we conclude that we are in the rapidly-rotating regime (see Fig. \ref{fig:TurbVisc}) in the cases of the Earth, Jupiter, Saturn, Uranus and Neptune. Therefore, the action of the Coriolis acceleration on the convective turbulent friction cannot be neglected that will strengthen the importance of tidal inertial waves and of their resonant viscous dissipation with corresponding numerous and strong resonances (see e.g. Fig. \ref{fig:spectra}). Note that such a result is particularly important for our understanding of tidal dissipation in the deep convective envelope of giant planets in our Solar System \citep[][]{OL2004,GMR2014} for which we obtained new constrains thanks to high precision astrometry \citep[][]{Laineyetal2009,Laineyetal2012,Laineyetal2016}.\\ 

We can thus conclude that in most of the astrophysical situations relevant for the evolution of planetary systems, the action of rotation on the turbulent friction applied by convection on linear tidal waves must be taken into account. Moreover, it demonstrates that {\it it is necessary to build a coupled modeling of the tidal evolution of planetary systems with the rotational evolution of their components, rotation being a key parameter for the amplitude and the frequency dependence of tidal dissipation in their interiors}.

\section{Conclusions and perspectives}

Using results obtained by \cite{Barkeretal2014} on the scaling of velocities and length scales in rotating turbulent convection zones, we proposed a new prescription for the eddy-viscosity coefficient, which allows us to describe tidal friction in such regions. In their work, \cite{Barkeretal2014} confirmed scalings for convective velocities and length scales as a function of rotation in the rapidly rotating regime that have been first derived by \cite{Stevenson1979} using mixing-length theory. Using his results, we derive a new prescription for the turbulent friction, which takes into account the action of the Coriolis acceleration on the convective flows. This allows us to generalize previous studies where the action of rotation on linear tidal flows was accounted for while its impact on the turbulent friction applied on them by convection was ignored. We demonstrated that the eddy-viscosity may be decreased by several ordres of magnitude in the rapidly rotating regime. It leads to a deep modification of the tidal dissipation frequency spectrum. On one hand, its background, that corresponds to the so-called equilibrium/non-wave like tide, is decreased because it scales as $E\propto\nu_{\rm T}$. On the other hand, resonances of tidal inertial waves ({\it i.e.} the dynamical tide) become more and more numerous, higher and sharper since their number, width at half-height, height, and sharpness respectively scales as $E^{-1/2}$, $E$, $E^{-1}$, and $E^{-2}$ \citep{ADMLP2015}. In this framework, we demonstrated thanks to {new} grids of rotating low-mass stars and values of convective Rossby numbers in planetary interiors \citep{SchubertSoderlund2011,Soderlundetal2013}, that this modification of the turbulent friction is important for low-mass stars along their PMS {and at the beginning of MS during which they are rapidly rotating} and in rapidly rotating planets as the Earth and giant gaseous/icy planets. {Because of the radial variation of the effects of rotation on the friction, they may be stronger for tidal inertial waves that propagate in the whole convective zone than for the equilibrium tide which has a higher amplitude in surface regions.} As demonstrated by \cite{Aetal2014}, such a behavior must be taken into account in the study of the tidal evolution of star-planet and planet-moon systems. Indeed, the angular velocity of their components vary along time because of structural modifications and of applied (tidal and electromagnetic) torques. These torques are themselves complex functions of rotation \citep[this work;][]{Mathis2015,Mattetal2012,Revilleetal2015,Mattetal2015}. It is thus necessary to have a completely coupled treatment of the tidal evolution of planetary systems and multiple stars and of the rotational evolution of their components with a coherent treatment of the variations of tidal flows and of their dissipation as a function of rotation.

In our work, the non-linear wave-wave interactions and those between tidal and convective flows have been ignored \citep[e.g.][]{Senetal2012,Sen2013,Barkeretal2013,Favieretal2014}. Moreover, stratified convection, with intermediate stably stratified diffusive layers can take place in giant planets because of double-diffusive instabilities \citep{LeconteChabrier2012}. Finally, stellar and planetary convective regions are differentially rotating and magnetized \citep[][]{BR2013,GBMR2016,BL2014}. In a near future, these four aspects of the problem will be examined carefully to improve our knowledge of tidal friction in stars and planets.

\begin{acknowledgements} 
The authors thank the referee for his positive and constructive report which allowed us to improve the article. The authors acknowledge funding by the European Research Council through ERC grant SPIRE 647383. This work was also supported by the Programme National de Plan\'etologie (CNRS/INSU), the GRAM specific action (CNRS/INSU-INP, CNES), the Axe Etoile of the Paris Observatory Scientific Council and the International Space Institute (ISSI team ENCELADE 2.0).
\end{acknowledgements}

\bibliographystyle{aa}  
\bibliography{Metalref} 

\end{document}